\documentclass[11pt,twoside]{article}
\usepackage{graphicx}
\usepackage{amsmath}
\usepackage{amssymb}
\usepackage{cite}
\usepackage{mathrsfs}

\begin{document}

\newcommand{\pst}{\hspace*{1.5em}}

\newcommand{\rigmark}{\em Journal of Russian Laser Research}
\newcommand{\lemark}{\em Volume 31, Number 2, 2010}

\newcommand{\be}{\begin{equation}}
\newcommand{\ee}{\end{equation}}
\newcommand{\bm}{\boldmath}
\newcommand{\ds}{\displaystyle}
\newcommand{\bea}{\begin{eqnarray}}
\newcommand{\eea}{\end{eqnarray}}
\newcommand{\ba}{\begin{array}}
\newcommand{\ea}{\end{array}}
\newcommand{\arcsinh}{\mathop{\rm arcsinh}\nolimits}
\newcommand{\arctanh}{\mathop{\rm arctanh}\nolimits}
\newcommand{\bc}{\begin{center}}
\newcommand{\ec}{\end{center}}

\thispagestyle{plain}

\label{sh}


\begin{center} {\Large \bf
\begin{tabular}{c}
SYMMETRIC INFORMATIONALLY COMPLETE\\[-1mm]
POSITIVE OPERATOR VALUED MEASURE\\[-1mm]
AND PROBABILITY REPRESENTATION\\[-1mm]
OF QUANTUM MECHANICS{\footnote{} }
\end{tabular}
 } \end{center}

{\footnotetext{Partially presented at the Workshop ``Nonlinearity
and Coherence in Classical and Quantum Systems'' held at the
University ``Federico~II'' in Naples, Italy on December 4, 2009 in
honor of Prof. Margarita A. Man'ko in connection with her 70th
birthday.}


\begin{center} {\bf
Sergey N. Filippov$^{1}$ and Vladimir I. Man'ko$^{2}$ }
\end{center}


\begin{center}
{\it
$^{1}$Moscow Institute of Physics and Technology (State University)\\
Institutskii per. 9, Dolgoprudnyi, Moscow Region 141700, Russia\\
\smallskip
$^2$P.~N.~Lebedev Physical Institute, Russian Academy of Sciences\\
Leninskii Prospect 53, Moscow 119991, Russia}

\smallskip

E-mail: sergey.filippov@phystech.edu, manko@sci.lebedev.ru
\end{center}

\begin{abstract}\noindent
Symmetric informationally complete positive operator valued
measures (SIC-POVMs) are studied within the framework of the
probability representation of quantum mechanics. A SIC-POVM is
shown to be a special case of the probability representation. The
problem of SIC-POVM existence is formulated in terms of symbols of
operators associated with a star-product quantization scheme. We
show that SIC-POVMs (if they do exist) must obey general rules of
the star product, and, starting from this fact, we derive new
relations on SIC-projectors. The case of qubits is considered in
detail, in particular, the relation between the SIC probability
representation and other probability representations is established,
the connection with mutually unbiased bases is discussed,
and comments to the Lie algebraic structure of SIC-POVMs are presented.
\end{abstract}


\section{\label{introduction} Introduction}
\pst
The properties of light beams in fibers~\cite{Rita1,Rita1a,Rita1b},
analytic signals~\cite{Rita2,Rita2a}, and quantum
systems~\cite{Rita3,Rita3a,Rita3b,Rita3c,FP}
are extensively studied, in particular,
within the framework of tomographic-probability representation.

The probability representation of quantum mechanics was
introduced recently in
\cite{tombesi-manko,m-t-m-FondPhys}. According to this
representation, the notion of wave function~\cite{schrodinger}
and density matrix~\cite{landau,von-neumann} can be replaced by
the notion of a fair probability distribution which determines the
quantum state. Indeed, the density operator (and all its other
phase-space representations like the Wigner function~\cite{wigner},
Husimi $Q$-function~\cite{husimi}, and Sudarshan--Glauber
$P$-function~\cite{sudarshan,glauber} for continuous degrees of
freedom) is related to the probability distribution with an integral
transform like the Radon transform~\cite{radon} (see also
\cite{gelfand}).

The probability representation is constructed
also for discrete spin variables~\cite{dodonovPLA,oman'ko-jetp}
and developed in \cite{serg-spin,serg-inverse-spin,serg-distances}
(see also the recent review~\cite{ibort}). From this point of
view, the probability representation of quantum mechanics is
completely equivalent to the other ones. On the other hand, this
representation has some new and unexpected aspects, for instance,
quantum states and transitions between them are described by
positive probabilities and positive transition probabilities,
respectively, instead of complex wave functions and complex
transition amplitudes inherent in the conventional formulation of
quantum mechanics. Thus, one can say that, in the probability
representation, the picture of quantum processes is similar to the
picture of classical processes in classical statistical mechanics,
where all the transitions are associated with transition
probabilities obeying to classical kinetic equations.

In the
probability representation, the quantum evolution equations, e.g.,
both Schr\"{o}dinger and von Neumann equations, can also be presented
in the form of classical-like kinetic equations for evolving
probability distributions.

There exist several different kinds of
the probability distributions, which are usually called tomographic
distributions or tomograms of quantum states. We can point out
optical tomograms~\cite{berber,vogel}, symplectic tomograms~\cite{mancini95},
spin tomograms~\cite{dodonovPLA,oman'ko-jetp},
photon-number tomograms~\cite{mancini,banaszek,wvogel}, and their
recent generalizations~\cite{asorey,asorey-arxiv}. In all the
tomographic pictures, the tomographic probability is a primary
concept of the quantum state. It is worth noting that the
challenging idea of trying to use a probability as a concept of the
quantum state was expressed in many earlier papers (see, e.g.,
\cite{ali-prugovecki-1,ali-prugovecki-2,ali-prugovecki-3,bush-lahti-FounPhys,
bush-lahti,bush-lahti-book,stulpe,stulpe-article,stulpe-book,kiukas}
and references therein), where the concept of informational
completeness was proposed. In spin tomography, Amiet and
Weigert~\cite{amiet-weigert-JPA} suggested an approach to the density
matrix reconstruction by using measurable probability
distributions and developed earlier results~\cite{newton}.

There exists a special way of describing quantum states in
finite-dimensional Hilbert spaces. This approach is initiated in
\cite{caves-sic,caves,renes}, developed substantially in
\cite{fuchs-2010}, and is known as symmetrical informationally
complete (SIC) approach. According to this viewpoint of quantum
mechanics (see, e.g, the recent concise review~\cite{fuchs-perimeter}),
quantum states are associated with probabilities connected with
a specific basis in the Hilbert space; with this basis being composed
of so-called SIC projectors.

Moreover, in the SIC approach, the probability distributions
describing the state density matrix contain no redundant
information, i.e., the number of probabilities is minimum possible
for reconstructing the density-matrix elements.

The main aim of our work
is to show that the SIC approach is equivalent to all other
available probability representations of quantum states in
finite-dimensional Hilbert spaces and connect the SIC approach
with the star-product~\cite{stratonovich}
formulation~\cite{oman'ko-JPA,oman'ko-vitale,serg-chebyshev,serg-PhysScr}
of the probability representation of quantum mechanics. In this paper,
we also point out a controversial disadvantage of the
SIC approach. The matter is that although the SIC representation
of quantum states is based on nonnegative probabilities summing to
unity, i.e., correct probability distributions from the
mathematical point of view, the SIC representation lacks for a
good physical interpretation of these probabilities. Namely, this
representation does not give a direct answer to the question: What
is the \textit{physical} quantity which can be measured
experimentally and gives rise to the probability distribution
involved?

The paper is organized as follows.

In Sec.~\ref{section-generic-star-product-scheme}, we review
generic star-product scheme of quantum mechanics following
\cite{oman'ko-JPA,oman'ko-vitale} and outline briefly a tomography
of spin states following
\cite{dodonovPLA,oman'ko-jetp,serg-spin,serg-inverse-spin}. In
Sec.~\ref{section-SIC}, the SIC approach is considered within the
framework of star-product scheme. In particular, the problem of
existence is formulated, an approach to its solution is discussed,
and a simple geometrical structure is presented. In
Sec.~\ref{section-sic-star-product}, we consider the star-product
scheme based on SIC projectors (the main goal of our work) and
derive new relations for these projectors. In
Sec.~\ref{section-Qubit}, the results obtained are applied to
qubits and discussed in detail. Section \ref{section-Lie} provides
some comments to the Lie algebraic structure of SIC projectors.
Finally, in Sec.~\ref{section-conclusions}, conclusions are
presented.

\section{\label{section-generic-star-product-scheme} Generic Star-Product Quantization Scheme}
\pst In this section, we are going to familiarize the reader with
a general structure of star-product quantization schemes to be
used extensively in subsequent sections. For the sake of
simplicity and brevity, we will restrict a mathematical rigor of
the development and omit the proofs that the definitions below are
introduced correctly. The good point is that only
finite-dimensional spaces will be focused on lately, so this
problem of rigor is less important than in the case of infinite
dimensions.

Let us consider a Hilbert space $\mathscr{H}$ and an operator
$\hat{A}$ acting on it. Then, such an operator can be
alternatively described by the following function $f_{A}({\bf x})$
of a set of variables ${\bf x}$:
\begin{equation}
\label{symbol} f_{A}({\bf x}) = {\rm Tr} \big[ \hat{A}\hat{U}({\bf
x}) \big],
\end{equation}
\noindent where $\hat{U}({\bf x})$ is a dequantizer operator~\cite{oman'ko-JPA}.
The function $f_{A}({\bf x})$ is often referred as a symbol of operator
$\hat{A}$. Once symbol $f_{A}({\bf x})$ is given, it is possible to find
an explicit form of the operator $\hat{A}$, making use of the quantizer operator
$\hat{D}({\bf x})$. Namely, the operator $\hat{A}$ reads
\begin{equation}
\label{A-from-symbol} \hat{A} = \int f_{A}({\bf x}) \hat{D}({\bf
x}) d{\bf x},
\end{equation}
\noindent where the set of variables ${\bf x}$ as well as the
integration $\int d{\bf x}$ depends on a system under study.
Obviously, the dequantizer and quantizer operators have different
explicit forms in different ${\bf x}$ representations. In
particular, as far as a spin-$j$ system is concerned, one can
alternatively utilize the following sets:
\begin{itemize}
\item ${\bf x}=(m,{\bf n})$, where $m$ is a spin projection on the
direction in space, ${\bf n}$, determined by a point
$(\cos\varphi\sin\theta,\sin\varphi\sin\theta,\cos\theta)$ on the
unit sphere $S^2$. In this case, the variable ${\bf n}$ is
continuous ($\varphi\in[0,2\pi]$, $\theta\in[0,\pi]$) and the
variable $m$ is discrete ($m=-j,-j+1,\ldots,j$). The integration
$\int d{\bf x}$ reduces to the summation
$\sum_{m=-j}^{j}(4\pi)^{-1}\int_{S^2}d{\bf n}$. The dequantizer
$\hat{U}(m,{\bf n})$ is introduced, and the quantizer
$\hat{D}(m,{\bf n})$ is found in the implicit form for such a
parametrization in \cite{dodonovPLA,oman'ko-jetp}.
Here, we will write both the dequantizer and quantizer in the
form developed in \cite{serg-spin}
\begin{equation}
\label{dequant-orth} \hat{U}(m,{\bf n}) = \sum_{L=0}^{2j}
f_L^{(j)}(m) \hat{S}_{L}^{(j)}({\bf n}), \qquad \hat{D}(m,{\bf n})
= \sum_{L=0}^{2j} (2L+1) f_L^{(j)}(m) \hat{S}_{L}^{(j)}({\bf
n}),\qquad
\end{equation}
\noindent with the coefficient $f_L^{(j)}(m)$ and the operator
$\hat{S}_L^{(j)}({\bf n})$ being expressed through the discrete
Chebyshev polynomial $t_n(x,N)$~\cite{nikiforov-suslov-uvarov} as
follows:
\begin{equation}
\label{discrete-Chebyshev} f_L^{(j)}(m) = \frac{1}{d_L}
t_L(j+m,2j+1), \qquad \hat{S}_L^{(j)}({\bf n}) = \left.
f_L^{(j)}(m) \right|_{m \longrightarrow (\hat{\bf J}\cdot {\bf n})
},
\end{equation}
\noindent where spin projection $m=-j,-j+1,\ldots,j$ is a discrete
variable\footnote{To be accurate, the function $d_L^{-1}
t_L(j+m,2j+1)$ is defined for discrete values of variable
$m=-j,-j+1,\ldots,j$, but we will associate this function with an
interpolation polynomial $f_L^{(j)}(m)$ of the lowest degree $L$.
Thus, the function $f_L^{(j)}(m)$ can be considered as a function of
continuous variable $m$, and the definition of operator function
$\hat{S}_L^{(j)}({\bf n}) = f_L^{(j)}\left( (\hat{\bf J}\cdot {\bf
n}) \right)$ is correct. For example,
$f_0^{(1/2)}(m)=\frac{1}{\sqrt{2}}$ and $f_1^{(1/2)}(m)= \sqrt{2} m$.
Consequently, $\hat{S}_0^{(1/2)}({\bf n})=
\frac{1}{\sqrt{2}}\hat{I}$ and $\hat{S}_1^{(1/2)}({\bf n})=
\sqrt{2} (\hat{\bf J}\cdot {\bf n})$, where $\hat{I}$ is, in
general, the $(2j+1)\times (2j+1)$ identity operator.} and the
normalization factor $d_L$ is $d_L=
\sqrt{\frac{(2j+L+1)!}{(2L+1)(2j-L)!}}$. Here, in passing we also
introduce a set of angular momentum operators $\hat{{\bf
J}}=(\hat{J}_x,\hat{J}_y,\hat{J}_z)$.
\item ${\bf x}=(m,u)$, where
$u$ is a general unitary rotation $u \in SU(N)$. This quantization
scheme is similar to the previous one and is considered in detail
in \cite{man'ko-sudarshan}. Note that a unitary spin tomographic
symbol $f_{A}(m,u)$ boils down to the spin tomographic symbol
$f_{A}(m,{\bf n})$ in the case of $u\in SU(2)$ representation.
\item
${\bf x}=(m,{\bf n}_k)$, with $\{{\bf n}_k\}_{k=1}^{4j+1}$ being a
finite set of directions ${\bf n}_k \in S^2$. If this is the case,
the integration $\int d{\bf x}$ implies the summation $\sum_{m=-j}^{j}
\sum_{k=1}^{4j+1}$. Following \cite{serg-inverse-spin},
the dequantizer and quantizer operators are
\begin{eqnarray}
&& \label{dequant-FNR} \hat{U}(m,{\bf n}_k) = (4j+1)^{-1}
f_L^{(j)}(m)
\hat{S}_L^{(j)}({\bf n}_k), \\
&& \label{quant-FNR} \hat{D}(m,k) = (4j+1) \sum\limits_{L:~
(k-1)/2 \le L\le 2j}
f_L^{(j)}(m)\sum_{k'=1}^{2L+1}\|\mathscr{M}^{-1}(L)\|_{kk'}
\hat{S}_L^{(j)}({\bf n}_{k'}), \qquad\qquad
\end{eqnarray}
where the $(2L+1)\times(2L+1)$ matrix $\|\mathscr{M}(L)\|$ is
readily expressed by virtue of the Legendre polynomial $P_l(x)$,
namely, its matrix elements read
\begin{equation}
\label{M-matrix} \|\mathscr{M}(L)\|_{kk'} = {\rm Tr}
\big[\hat{S}_L^{(j)}({\bf n}_k)\hat{S}_L^{(j)}({\bf n}_{k'}) \big]
= P_{L}({\bf n}_{k}\cdot{\bf n}_{k'}).
\end{equation}
\end{itemize}

Now, if we substitute the density operator $\hat{\rho}$ of a
spin-$j$ system for $\hat{A}$ in definition (\ref{symbol}) and
choose one of the quantization schemes above, the corresponding
symbol $w({\bf x}) \equiv f_{\rho}({\bf x})$ is a fair probability
distribution function also known as a tomogram -- spin tomogram
$w(m,{\bf n})$, unitary spin tomogram $w(m,u)$, and spin tomogram
with a finite number of rotations $w(m,{\bf n}_k)$ (spin-FNR
tomogram), respectively.

It is worth mentioning, that the function $\mathfrak{D}({\bf x},{\bf
x}') = {\rm Tr} \big[ \hat{U}({\bf x}) \hat{D}({\bf x}') \big]$
has a sense of delta-function on symbols $f_{A}({\bf x})$. This
means that
\begin{equation}
\label{star-product-delta} f_{A}({\bf x}) = \int \mathfrak{D}({\bf
x},{\bf x}') f_{A}({\bf x}') d{\bf x}'.
\end{equation}

\subsection{\label{subsection-Star-product} Star Product}
\pst Let us now consider a symbol $f_{AB}({\bf x})$ of the product
of two operators $\hat{A}$ and $\hat{B}$ acting on $\mathscr{H}$.
The symbol $f_{AB}({\bf x})$ is referred as the star product of
symbols $f_{A}({\bf x}_1)$ and $f_{B}({\bf x}_2)$ and is obtained
by the formula
\begin{equation}
f_{AB}({\bf x}) \equiv (f_{A} \star f_{B}) ({\bf x}) = \int
f_{A}({\bf x}_1) f_{B}({\bf x}_2) K({\bf x}_1,{\bf x}_2,{\bf x})
d{\bf x}_1 d{\bf x}_2,
\end{equation}
\noindent where the star-product kernel $ K({\bf x}_1,{\bf
x}_2,{\bf x})$ reads
\begin{equation}
\label{star-product-kernel} K({\bf x}_1,{\bf x}_2,{\bf x}) = {\rm
Tr} \big[ \hat{D}({\bf x}_1) \hat{D}({\bf x}_2) \hat{U}({\bf x})
\big].
\end{equation}
\noindent It is easily seen that the star product is associative
but not necessarily commutative. The associativity property has
important consequences. In particular, the star-product kernel
$K^{(N)}({\bf x}_1,\ldots,{\bf x}_N,{\bf x})$ of an arbitrary
number $N$ of symbols is expressed through the kernel
(\ref{star-product-kernel}). For example, in the case of three
operators $\hat{A}$, $\hat{B}$, and $\hat{C}$, we have
\begin{equation}
f_{ABC}({\bf x}) = (f_{A} \star f_{B} \star f_{C}) ({\bf x}) =
\big((f_{A} \star f_{B}) \star f_{C}\big) ({\bf x}) = \big(f_{A}
\star (f_{B} \star f_{C}) \big) ({\bf x}), \qquad\qquad
\end{equation}
\noindent from which it follows that
\begin{equation}
\label{kernel-3} K^{(3)}({\bf x}_1,{\bf x}_2,{\bf x}_3,{\bf x}) =
\int K({\bf x}_1,{\bf x}_2,{\bf y}) K({\bf y},{\bf x}_3,{\bf x})
d{\bf y} = \int K({\bf x}_1,{\bf y},{\bf x}) K({\bf x}_2,{\bf
x}_3,{\bf y}) d{\bf y}. \qquad
\end{equation}
\noindent Similarly, in the case of four operators, we obtain
\begin{eqnarray}
\label{kernel-4} && K^{(4)}({\bf x}_1,{\bf x}_2,{\bf x}_3,{\bf
x}_4,{\bf x}) = \int K({\bf x}_1,{\bf x}_2,{\bf y}) K({\bf y},{\bf
x}_3,{\bf z}) K({\bf
z},{\bf x}_4,{\bf x}) d{\bf y} d{\bf z} \nonumber\\
&& = \int K({\bf x}_1,{\bf x}_2,{\bf y}) K({\bf y},{\bf z},{\bf
x}) K({\bf x}_3,{\bf x}_4,{\bf z}) d{\bf y} d{\bf z} = \int K({\bf
x}_1,{\bf y},{\bf z}) K({\bf x}_2,{\bf x}_3,{\bf
y}) K({\bf z},{\bf x}_4,{\bf x}) d{\bf y} d{\bf z} \nonumber\\
&& = \int K({\bf x}_1,{\bf y},{\bf x}) K({\bf x}_2,{\bf x}_3,{\bf
z}) K({\bf z},{\bf x}_4,{\bf y}) d{\bf y} d{\bf z} = \int K({\bf
x}_1,{\bf y},{\bf x}) K({\bf x}_2,{\bf z},{\bf y}) K({\bf
x}_3,{\bf x}_4,{\bf z}) d{\bf y} d{\bf z}. \qquad\qquad
\end{eqnarray}
\noindent Equalities (\ref{kernel-3})--(\ref{kernel-4}) impose
limitations on the star-product kernel (\ref{star-product-kernel})
and will be considered with regards to the SIC star-product scheme
in Sec.~\ref{section-sic-star-product}.

\subsection{\label{subsection-dual-symbols}Dual Symbols}
\pst The quantization scheme (\ref{symbol})--(\ref{A-from-symbol})
has a dual one defined by the relations~\cite{oman'ko-vitale}
\begin{equation}
\label{dual} f_{A}^{\rm dual}({\bf x}) = {\rm Tr} \big[
\hat{A}\hat{D}({\bf x}) \big], \qquad \hat{A} = \int f_{A}^{\rm
dual}({\bf x}) \hat{U}({\bf x}) d{\bf x}.
\end{equation}
\noindent Arguing as above, we obtain the star-product kernel of
dual symbols in the form
\begin{equation}
\label{star-product-kernel-dual} K^{\rm dual}({\bf x}_1,{\bf
x}_2,{\bf x}) = {\rm Tr} \big[ \hat{U}({\bf x}_1) \hat{U}({\bf
x}_2) \hat{D}({\bf x}) \big].
\end{equation}
\noindent This kernel exhibits the same general properties as the
star-product kernel (\ref{star-product-kernel}); in particular,
the relations analogues to (\ref{kernel-3})--(\ref{kernel-4}) take
place (one should merely replace $K$ by $K^{\rm dual}$).

\section{\label{section-SIC}SIC-POVMs}
\pst Recently, much attention has been paid to a highly symmetric
informationally complete positive operator valued measure
(SIC-POVM) in $d$-dimensional Hilbert space $\mathscr{H}_d$ (see,
e.g., the review~\cite{fuchs-2010}). The existence of SIC-POVMs in
any finite dimension still remains an unsolved problem, though
astonishing results are obtained in both analytical and numerical
investigations, namely, the existence is effectively demonstrated
in dimensions $d\le 67$ \cite{scott-grassl}. The core of any
SIC-POVM is a set of $d^2$ rank-1 projectors $\hat{\Pi}_i =
|\psi_i\rangle\langle\psi_i|$ acting on $\mathscr{H}_d$ and
satisfying the condition
\begin{equation}
\label{scalarproduct} {\rm Tr}\big[\hat{\Pi}_i\hat{\Pi}_j\big] =
\left| \langle\psi_i|\psi_j\rangle \right|^2 =
\frac{d\delta_{ij}+1}{d+1},
\end{equation}
\noindent where $\delta_{ij}$ is the Kronecker delta-symbol.

\subsection{\label{subsection-SIC-representation} SIC Representation of Quantum States}
\pst The SIC representation of quantum states \cite{ericsson} is
based on the idea that a quantum state, usually described by the
density operator $\hat{\rho}$, is also fully determined by $d^2$
probabilities $p_i$. The set of probabilities
$\{p_i\}_{i=1}^{d^2}$ and the density-operator reconstruction read
\begin{equation}
\label{SIC-scheme} p_i = \frac{1}{d} {\rm Tr} \big[ \hat{\rho}
\hat{\Pi}_i \big], \qquad \hat{\rho} = (d+1)\sum_{i=1}^{d^2} p_i
\hat{\Pi}_i - \hat{I}.
\end{equation}
\noindent In accordance with the SIC representation, every quantum
state can be represented as a set of probabilities
$\{p_i\}_{i=1}^{d^2}$ in the simplex of all probability vectors
with $d^2$ components.

It is worthwhile clarifying a possible drawback of the SIC
representation (and, indeed, of many other POVMs). Following
general ideas of the POVM construction, the set of probabilities
$\{p_i\}_{i=1}^{d^2}$ is often referred as the probability
distribution since $0 \le p_i \le 1$ for all $i=1,\ldots,d^2$ and
$\sum_{i=1}^{d^2}p_i = 1$. Although such a treatment is correct
from the mathematical point of view; in physics, one also needs an
interpretation of this probability distribution. In fact, one
needs to associate the probabilities with the relative frequency
of outcomes of a physical quantity which can be measured
experimentally. Thus, the following conceptual problem arises
itself: What \textit{physical} quantity should one measure in
order to obtain the ``probability distribution"
$\{p_i\}_{i=1}^{d^2}$? This problem is analogous to that
concerning the probabilistic treatment of the Husimi $Q$-function
\cite{husimi}. Although $Q$-function is nonnegative and
normalized, it cannot be considered as a fair probability
distribution. This is because $Q$-function does not have sense of
a joint probability-distribution function in the phase space. As
far as the spin tomography~\cite{dodonovPLA,oman'ko-jetp}, the
unitary spin tomography~\cite{man'ko-sudarshan}, and the spin
tomography with a finite number of
rotations~\cite{serg-inverse-spin} are concerned, such a problem
of the physical meaning does not arise since the density operator
is related to a special distribution functions of physical
observables (spin projection). Nevertheless, we must admit that
formulas developed in
\cite{dodonovPLA,oman'ko-jetp,man'ko-sudarshan,serg-inverse-spin}
are similar to that developed in the SIC representation of quantum
states.

To eliminate this (controversial) drawback, one can think of
probabilities~(\ref{SIC-scheme}) as a part of a physical
probability distribution based on the idea of the inverse spin-$s$
portrait~\cite{serg-inverse-spin}. Namely, for a spin-$j$ system
($d=2j+1$), all the vectors $|\psi_i\rangle \in \mathbb{C}^d$,
$i=1,\ldots,d^2$ can be expressed through the highest-projection
eigenstate $|jj\rangle$ of the angular momentum operators
$\hat{J}_z$ and $\hat{\bf J}^2$ as follows: $|\psi_i\rangle =u_i
|jj\rangle$, where $\{u_i\}_{i=1}^{d^2}$ is a specific set of
$d$$\times$$d$ unitary matrices. Then the physical probability
distribution reads
\begin{equation}
\mathcal{P}(m,i) = \frac{1}{d^2} \langle jm | u_i^{\dag}
\hat{\rho} u_i |j m\rangle.
\end{equation}
\noindent The factor $1/d^2$ is assigned to a priori probability
to choose a rotation $u_i$ (labeled by a random quantity $i$) in
the Hilbert space $\mathscr{H}_d$, whereas $\langle jm |
u_i^{\dag} \hat{\rho} u_i |j m\rangle$ is the probability to
obtain the spin projection $m$ on the $z$ axis after rotation
$u_i$ in the Hilbert space is fulfilled. Note that
$\sum_{m=-j}^{j} \mathcal{P}(m,i) = 1/d^2$, $\sum_{i=1}^{d^2}
\mathcal{P}(j,i) = d$, and $\sum_{m=-j}^{j}\sum_{i=1}^{d^2}
\mathcal{P}(m,i) = 1$.
The relation to SIC probabilities~(\ref{SIC-scheme}) reads $p_i=\mathcal{P}(j,i)$.

Also, it is worth mentioning the relation ${\rm Tr}\big[
\hat{\rho} \hat{\Pi}_i \big] =
\langle\psi_i|\hat{\rho}|\psi_i\rangle$. This relation means that
the trace in formula (\ref{SIC-scheme}) can be interpreted not
only as the mean value of the observable $\hat{\Pi}_i$ in the
state $\hat{\rho}$, but also as a probability
$\langle\psi_i|\hat{\rho}|\psi_i\rangle$ which is nothing else but
the mean value of the operator $\hat{\rho}$ in the state
$|\psi_i\rangle$ (compare with the notion of expectation value
used in \cite{weigert-PRL}). For example, in the case of qubits
(spin-$1/2$ system) one can think of the
density matrix $\hat{\rho}_{\uparrow} = \left(%
\begin{array}{cc}
  1 & 0 \\
  0 & 0 \\
\end{array}%
\right)$ as an observable $A$ such that its outcomes read: $A=1$
if the measurement of spin projection $m$ on the $z$ axis results
in $m=+1/2$ and $A=0$, otherwise. In other words, the observable
$A$ is given by the operator $\hat{A} =
\frac{1}{2}(\hat{I}+\hat{\sigma}_z)$. Then the trace ${\rm
Tr}\big[ \hat{\rho}_{\uparrow} \hat{\Pi}_i \big] =
\langle\psi_i|\hat{A}|\psi_i\rangle$ is a mean value of the observable
$A$ in the state $|\psi_i\rangle$ or, equivalently, the
probability to obtain the outcome $A=1$ in the state
$|\psi_i\rangle$.

\subsection{\label{subsection-SIC-de-quantizer} SIC Dequantizer and SIC Quantizer}
\pst Comparing the definitions of the SIC scheme
(\ref{SIC-scheme}) with the star-product construction, we see that
the SIC-POVM effects $\hat{U}_i$ are then defined by $\hat{U}_i =
\frac{1}{d}\hat{\Pi}_i$ and represent themselves nothing else but
SIC dequantizers $\hat{U}({\bf x})$ depending on discrete variable
${\bf x} = i$, $i=1,\ldots,d^2$. Also, SIC dequantizers sum to
unity, i.e., $\sum_{i=1}^{d^2} \hat{U}_i = \hat{I}$. The existence
of a SIC quantizer $\hat{D}_i \equiv \hat{D}({\bf x}=i)$ is
equivalent to the POVM being informationally complete. Comparison
of the SIC approach (\ref{SIC-scheme}) with formula
(\ref{A-from-symbol}) yields that the SIC quantizer $\hat{D}_i$ is
expressed in terms of projectors $\{\hat{\Pi}_k\}_{k=1}^{d^2}$
(see, e.g., \cite{caves-sic})
\begin{equation}
\hat{D}_i = (d+1)\hat{\Pi}_i - \hat{I}.
\end{equation}
Taking into account the positivity of projectors $\hat{\Pi}_i \ge
0$, the SIC-symbol $f_{\rho i} = {\rm
Tr}\big[\hat{\rho}\hat{U}_i\big]$ of any density operator
$\hat{\rho}$ is obviously nonnegative and, consequently, defines a
fair probability distribution function (SIC tomogram) $w_i$
depending on discrete parameter $i=1,\ldots,d^2$. It is worth
noting that the SIC-quantization scheme is a particular case of
the general problem of mapping an abstract Hilbert space on the
set of tomograms (fair probability distributions) within the
framework of the probability picture of quantum mechanics. The
method for constructing such a tomographic setting is developed in
\cite{manko-sudarshan-vent}. The only restriction to such a
setting is the condition~(\ref{scalarproduct}) to be met, though
we must admit that the explicit form of projectors
$\{\hat{\Pi}_k\}_{k=1}^{d^2}$ is rather difficult to find in any
$d$-dimensional Hilbert space $\mathscr{H}_d$.

\subsection{\label{subsection-existence-problem}
Existence Problem in Terms of Symbols of Operators}
\pst Since any operator can be associated with a corresponding
symbol of the form (\ref{symbol}), let us reformulate the problem
of finding the SIC-POVM in terms of symbols $f({\bf x})$. To
start, a set of operators $\{\frac{1}{d}\hat{\Pi}_j\}_{j=1}^{d^2}$
forms the SIC-POVM iff for all $i=1,\ldots,d^2$ the following
conditions altogether are fulfilled:
\begin{eqnarray}
&& \label{hermicity} \hat{\Pi}_i^{\dag} = \hat{\Pi}_i,\\
&& \label{unity} {\rm Tr} \big[ \hat{\Pi}_i \big] = 1,\\
&& \label{positivity} \hat{\Pi}_i \ge 0, {\rm \quad i.e.,\quad} \langle\psi
|\hat{\Pi}_i|\psi\rangle \ge 0 {\rm ~for~all~} |\psi\rangle,\\
&& \label{projector} \hat{\Pi}_i^{2} = \hat{\Pi}_i,\\
&& \label{trace}{\rm Tr} \big[ \hat{\Pi}_i\hat{\Pi}_j \big] =
\frac{d\delta_{ij}+1}{d+1}.
\end{eqnarray}
\noindent It is shown in \cite{jones-linden} that the
normalization condition (\ref{unity}), the positivity condition
(\ref{positivity}), and the projectivity property
(\ref{projector}) can be unified for a Hermitian operator
$\hat{\Pi}_i$ in the form of a trace equalities
\begin{equation}
(\ref{unity}) \wedge (\ref{positivity}) \wedge (\ref{projector})
\quad \Longleftrightarrow \quad {\rm Tr} \big[ \hat{\Pi}_i \big]
 ={\rm Tr} \big[ \hat{\Pi}_i^{2} \big] = {\rm Tr} \big[ \hat{\Pi}_i^{3} \big]
 = 1 \quad ~{\rm if}~ (\ref{hermicity}) ~{\rm is ~ true}.
\end{equation}

Now, by $f_{\Pi_i}({\bf x}) \equiv {\rm Tr}\big[ \hat{\Pi}_i
\hat{U}({\bf x}) \big]$ denote the symbol (\ref{symbol}) of the
operator $\hat{\Pi}_i$ in some star-product quantization scheme
defined by dequanizer $\hat{U}({\bf x})$ and quantizer
$\hat{D}({\bf x})$. Then, symbols $\{f_{\Pi_i}({\bf
x})\}_{i=1}^{d^2}$ correspond to SIC projectors iff
\begin{eqnarray}
&& \label{hermicity-symbols} f_{\Pi_i}({\bf x}) = \int f_{\Pi_i}^{\ast}({\bf y}) {\rm Tr} \big[ \hat{U}({\bf x}) \hat{D}^{\dag}({\bf y}) \big] d{\bf y},\\
&& \label{trace-1-symbols} \int f_{\Pi_i}({\bf x}) {\rm Tr} \big[ \hat{D}({\bf x}) \big] d{\bf x} = 1,\\
&& \label{trace-2-symbols} \int f_{\Pi_i}({\bf x}) f_{\Pi_j}({\bf
y}) {\rm Tr} \big[ \hat{D}({\bf x})
\hat{D}({\bf y}) \big] d{\bf x} d{\bf y} = \frac{d\delta_{ij}+1}{d+1},\\
&& \label{trace-3-symbols} \int f_{\Pi_i}({\bf x}) f_{\Pi_i}({\bf
y}) f_{\Pi_i}({\bf z}) {\rm Tr} \big[ \hat{D}({\bf x})
\hat{D}({\bf y}) \hat{D}({\bf z}) \big] d{\bf x} d{\bf y} d{\bf z}
= 1.
\end{eqnarray}

Thus, in a particular star-product scheme, the problem of seeking
the SIC projectors transforms into the problem of seeking the
corresponding symbols.

\subsection{\label{subsection-search}
Example: Search of SIC Projectors in a Concrete Quantization Scheme}
\pst To demonstrate such an approach to the SIC existence problem,
let us consider the following star-product quantization scheme:
\begin{eqnarray}
&& {\bf x} = (L,{\bf n}_k), \qquad k=1,2,\ldots,2L+1, \qquad
L=0,1,\ldots,d-1, \qquad \int d{\bf x} = \sum_{L=0}^{d-1} \sum_{k=1}^{2L+1},  \\
&& \hat{U}({\bf x}) \equiv \hat{U}(L,{\bf n}_k) = \hat{S}_{L}({\bf
n}_k), \qquad \hat{D}({\bf x}) \equiv \hat{D}(L,k) =
\sum_{k'=1}^{2L+1} \| \mathscr{M}^{-1}(L) \|_{kk'}
\hat{S}_{L}({\bf n}_{k'}),
\end{eqnarray}
\noindent where the operator $\hat{S}_L({\bf n}_k)$ and the matrix
$\| \mathscr{M}(L) \|$ are defined by formulas
(\ref{discrete-Chebyshev}) and (\ref{M-matrix}), respectively. In
this quantization scheme, both dequantizer and quantizer are
Hermitian. Then, in view of relation (\ref{star-product-delta}), it
is not hard to see that the hermicity condition
(\ref{hermicity-symbols}) is fulfilled whenever the corresponding
symbols are real, i.e., $f_{\Pi_i}({\bf x}) \equiv f_{\Pi_i}(L,{\bf
n}_k) \in \mathbb{R}$ for all $i=1,\ldots,d^2$. Proceeding to the
necessary condition (\ref{trace-1-symbols}), we utilize specific
properties of the operator $\hat{S}_L({\bf n}_k)$ (see, e.g.,
\cite{serg-spin,serg-inverse-spin}), namely, ${\rm Tr}\big[
\hat{S}_{L}({\bf n}_k) \big] = \sqrt{d} \delta_{L0}$. Combining
this with the evident relation $\| \mathscr{M}(L=0) \| = 1$, we
obtain $(\ref{trace-1-symbols}) \Longleftrightarrow
f_{\Pi_i}(0,{\bf n}_1) = \frac{1}{\sqrt{d}}$. Further, to consider
properties (\ref{trace-1-symbols})--(\ref{trace-3-symbols}), it is
convenient to write symbols $f_{\Pi_i}(L,{\bf n}_k)$ in the form
of vectors ${\bf f}_i$ for each $i=1,\ldots,d^2$ and collect them
into the following $d^2$$\times$$d^2$ matrix $\|F\|$:
\begin{equation}
{\bf f}_i =
\left(%
\begin{array}{c}
  f_{\Pi_i}(0,{\bf n}_1) \\
  \hline
  f_{\Pi_i}(1,{\bf n}_1) \\
  f_{\Pi_i}(1,{\bf n}_2) \\
  f_{\Pi_i}(1,{\bf n}_3) \\
  \hline
  \vdots \\
  \hline
  f_{\Pi_i}(d-1,{\bf n}_1) \\
  \vdots \\
  f_{\Pi_i}(d-1,{\bf n}_{2d-1}) \\
\end{array}%
\right), \qquad \|F\| = \left\|%
\begin{array}{cccc}
  {\bf f}_1 & {\bf f}_2 & \cdots & {\bf f}_{d^2} \\
\end{array}%
\right\|.
\end{equation}
\noindent Indeed, this is the matrix $\|F\|$ that determines
explicitly the set of SIC projectors
$\{\hat{\Pi}_i\}_{i=1}^{d^2}$. Indeed, stacking projectors
$\hat{\Pi}_i$ into the vector operator $\hat{\boldsymbol{\Pi}}$
and, in the same way, operators $\hat{D}(L,k)$ into the vector
operator $\hat{\bf D}$, we readily obtain
\begin{equation}
\label{Pi-vector} \hat{\boldsymbol{\Pi}} = \| F \|^{\rm T}
\hat{\bf D}.
\end{equation}
\noindent Now, requirement (\ref{trace-2-symbols}) can be
rewritten as follows:
\begin{equation}
\label{gram-elements} {\rm Tr} \big[ \hat{\Pi}_i \hat{\Pi}_j \big]
= \sum_{L=0}^{d-1}\sum_{k,k'=1}^{2L+1} \| \mathscr{M}^{-1}(L)
\|_{kk'} f_{\Pi_i}(L,k) f_{\Pi_j}(L,k') =
\frac{d\delta_{ij}+1}{d+1} \equiv \|\Gamma\|_{ij},
\end{equation}
\noindent or briefly in the form of the following matrix equation:
\begin{equation}
\label{F-M-F-Gamma} \| F \|^{\rm T} \| \mathfrak{M}^{-1} \| \|F\|
= \|\Gamma\|,
\end{equation}
\noindent where we introduced two new $d^2$$\times$$d^2$ matrices --
the Gram matrix $\|\Gamma\|$ with matrix elements~(\ref{gram-elements})
and the block-diagonal matrix $\|\mathfrak{M} \|$ defined through
\begin{equation}
\| \mathfrak{M} \| = \left(
\begin{array}{cccc}
\underset{1\times 1}{\mathscr{M}(L=0)} &\multicolumn{3}{|c}{}\\
\cline{1-2} \multicolumn{1}{c|}{} & \underset{3\times
3}{\mathscr{M}(L=1)} &\multicolumn{2}{|c}
{\raisebox{1.5ex}[0pt]{\parbox{12pt}{\Huge 0}}}\\
\cline{2-2} \multicolumn{2}{c}{}& \ddots &   \\
\cline{4-4} \multicolumn{3}{c|}
{\raisebox{1.5ex}[0pt]{\parbox{12pt}{\Huge 0}}} &
\underset{(2d-1)\times (2d-1)}{\mathscr{M}(L=d-1)}
\end{array}
 \right).
\end{equation}
\noindent The Gram matrix $\|\Gamma\|$ can be represented in the
form of the product $\|\mathscr{S}\|^{\rm T}\|\mathscr{S}\|$,
where $\|\mathscr{S}\|$ is a transition matrix to the orthonormal
basis~\cite{gantmacher}. The expansion
$\|\Gamma\|=\|\mathscr{S}\|^{\rm T}\|\mathscr{S}\|$ is not unique
and, in general, takes the form $\|\Gamma\|=\|\mathscr{S}\|^{\rm
T}\|{Q}_{\Gamma}\|^{\rm T}\|{Q}_{\Gamma}\|\|\mathscr{S}\|$, where
$\|Q_{\Gamma}\|$ is an arbitrary real orthogonal matrix. IN view of the
same argument, we obtain a similar expansion $\| \mathfrak{M} \| =
\|\mathfrak{S}\|^{\rm T}\|{Q}_{\mathfrak{M}}\|^{\rm
T}\|{Q}_{\mathfrak{M}}\|\|\mathfrak{S}\|$. We succeeded in finding the
explicit form of matrices $\|\mathfrak{S}\|$ and $\|\mathscr{S}\|$
in any dimension $d$. The matrix $\|\mathfrak{S}\|$ has the
block-diagonal form
\begin{equation} \label{S-frak} \| \mathfrak{S} \| = \left(
\begin{array}{cccc}
\underset{1\times 1}{\mathit{\Sigma}(L=0)} &\multicolumn{3}{|c}{}\\
\cline{1-2} \multicolumn{1}{c|}{} & \underset{3\times
3}{\mathit{\Sigma}(L=1)} &\multicolumn{2}{|c}
{\raisebox{1.5ex}[0pt]{\parbox{12pt}{\Huge 0}}}\\
\cline{2-2} \multicolumn{2}{c}{}& \ddots &   \\
\cline{4-4} \multicolumn{3}{c|}
{\raisebox{1.5ex}[0pt]{\parbox{12pt}{\Huge 0}}} &
\underset{(2d-1)\times (2d-1)}{\mathit{\Sigma}(L=d-1)}
\end{array}
 \right),
\end{equation}
\noindent with each block being expressed through associate
Legendre polynomials $P_l^{(m)}(x)$ and vectors ${\bf n}_j =
(\cos\varphi_j\sin\theta_j,\sin\varphi_j\sin\theta_j,\cos\theta_j)$
as follows:
\begin{equation}
\label{Sigma-blocks}
  \qquad \|\mathit{\Sigma}(L)\|_{ij} = \left\{ \begin{array}{ll}
   P_L^{(0)} (\cos \theta_j) & {\rm if}~ i=1, \\
   \sqrt{\frac{2(L-m)!}{(L+m)!}} P_L^{(m)} (\cos \theta_j) \cos m \varphi_j & {\rm if} ~ i=2m, ~ m=1,2,\ldots , \\
   \sqrt{\frac{2(L-m)!}{(L+m)!}} P_L^{(m)} (\cos \theta_j) \sin m \varphi_j & {\rm if} ~ i=2m+1, ~ m=1,2,\ldots  \\
 \end{array} \right.
\end{equation}
\noindent The explicit expression of matrix $\|\mathscr{S}\|$ in
any dimension $d$ of the Hilbert space reads
\begin{equation}
\label{S-scr}
\|\mathscr{S}\| = \left(%
\begin{array}{cccccc}
  \frac{1}{\sqrt{d}} & \frac{1}{\sqrt{d}} & \frac{1}{\sqrt{d}} & \frac{1}{\sqrt{d}} & \cdots & \frac{1}{\sqrt{d}} \\
  -\sqrt{\frac{d}{2(d+1)}} & \sqrt{\frac{d}{2(d+1)}} & 0 & 0 & \cdots & 0\\
  -\sqrt{\frac{d}{6(d+1)}} & -\sqrt{\frac{d}{6(d+1)}} & \sqrt{\frac{2d}{3(d+1)}} & 0 & \cdots & 0 \\
  -\sqrt{\frac{d}{12(d+1)}} & -\sqrt{\frac{d}{12(d+1)}} & -\sqrt{\frac{d}{12(d+1)}} & \sqrt{\frac{3d}{4(d+1)}} & \cdots & 0 \\
  \vdots & \vdots & \vdots & \vdots & \ddots & \vdots \\
  -\frac{1}{(d+1)\sqrt{d(d-1)}} & -\frac{1}{(d+1)\sqrt{d(d-1)}} & -\frac{1}{(d+1)\sqrt{d(d-1)}} & -\frac{1}{(d+1)\sqrt{d(d-1)}} & \cdots & \sqrt{\frac{d-1}{d}} \\
\end{array}%
\right),
\end{equation}
\noindent or in the most general form
\begin{equation}
\label{S-scr-general} \|\mathscr{S}\|_{1l}=\frac{1}{\sqrt{d}},
\qquad \|\mathscr{S}\|_{kl} = \left\{
\begin{array}{cl}
  -\sqrt{\frac{d}{k(k-1)(d+1)}} & {\rm if}~  l<k, \\
  \sqrt{\frac{(k-1)d}{k(d+1)}} & {\rm if}~  l=k, \\
  0 & {\rm if}~  l>k \\
\end{array} \right. \quad {\rm for ~ all}~ k=2,\ldots,d^2.
\qquad\quad
\end{equation}
\noindent It is worth mentioning that the condition
$\sum_{i=1}^{d^2}\hat{\Pi}_i=d\hat{I}$ reduces to the condition
$\sum_{l=1}^{d^2}\|\mathscr{S}\|_{kl}=\delta_{k1}d^{3/2}$ which is
obviously fulfilled.

Further, applying the obtained results to formula~(\ref{F-M-F-Gamma}) yields
\begin{eqnarray}
& \| F \|^{\rm T} \|\mathfrak{S}^{-1}\|\|{Q}_{\mathfrak{M}}\|^{\rm
T} \|{Q}_{\mathfrak{M}}\| \|\mathfrak{S}^{-1}\|^{\rm T}
 \|F\| = \|\mathscr{S}\|^{\rm
T}\|{Q}_{\Gamma}\|^{\rm T}\|{Q}_{\Gamma}\|\|\mathscr{S}\|,\\
& \|F\| = \|\mathfrak{S}\|^{\rm T} \|{Q}_{\mathfrak{M}}\|^{\rm T}
\|{Q}_{\Gamma}\| \|\mathscr{S}\|,
\end{eqnarray}
\noindent where we have taken into account the main property of
orthogonal matrices $\|Q\|^{-1}=\|Q\|^{\rm T}$. Finally, let
$\|Q\| = \|{Q}_{\mathfrak{M}}\|^{\rm T} \|{Q}_{\Gamma}\|$ be the
resulting orthogonal matrix, and $\hat{\bf S}$ be a vector operator
with components $\hat{S}_L({\bf n}_k)$, $k=1,\ldots,2L+1$,
$L=0,\ldots,d-1$, then formula (\ref{Pi-vector}) transforms into
\begin{equation}
\label{Pi-vector-found} \hat{\boldsymbol{\Pi}} = \| \mathscr{S}
\|^{\rm T} \| Q \|^{\rm T} \|\mathfrak{S}\| \hat{\bf D} = \|
\mathscr{S} \|^{\rm T} \| Q \|^{\rm T} \|\mathfrak{S}^{-1}\|^{\rm
T} \hat{\bf S},
\end{equation}
\noindent which expresses unknown SIC projectors $\hat{\Pi}_j$ in
terms of known matrices $\|\mathscr{S}\|$, $\|\mathfrak{S}\|$, and
operators $\hat{S}_L({\bf n}_k)$ given by formulas~(\ref{S-scr})--(\ref{S-scr-general}),
(\ref{S-frak})--(\ref{Sigma-blocks}), and
(\ref{discrete-Chebyshev}), respectively. The only unspecified
matrix is the orthogonal matrix $\|Q\|$. Now, if we recall the
trace condition $(\ref{trace-1-symbols}) \Longleftrightarrow
f_{\Pi_i}(0,{\bf n}_1) = \frac{1}{\sqrt{d}}$, we will readily see
that the $d^2$$\times$$d^2$ orthogonal matrix $\|Q\|$ must be
block-diagonal
\begin{equation}
\label{Q-block}
\|Q\|=\left(%
\begin{array}{c|ccc}
  1 & 0 & \cdots & 0\\
  \hline
  0 & \multicolumn{3}{c}{} \\
  \vdots & \multicolumn{3}{c}{\underset{\raisebox{0ex}[0pt]
  {\parbox{50pt}{\tiny $(d^2\!-\!1)\!\times\!(d^2\!-\!1)$}}}{\widetilde{Q}}}\\
  0 & \multicolumn{3}{c}{}\\
\end{array}%
\right).
\end{equation}

Though the matrix $\|Q\|$ remains precisely undetermined,
constructed in such a manner set of operators
$\{\hat{\Pi}_i\}_{i=1}^{d^2}$~(\ref{Pi-vector-found}) does satisfy
requirements (\ref{hermicity-symbols})--(\ref{trace-2-symbols})
for any chosen orthogonal matrix $\|Q\|$ of the form~(\ref{Q-block}).
Thus, we have constructed a set of Hermitian operators
$\{\hat{\Pi}_i\}_{i=1}^{d^2}$ such that ${\rm Tr}\big[
\hat{\Pi}_i \big] = {\rm Tr}\big[ \hat{\Pi}_i^2 \big] = 1$ and
${\rm Tr}\big[ \hat{\Pi}_i \hat{\Pi}_j \big] = \frac{1}{d+1}$ if
$i \ne j$.
Furthermore, the last but not least is the condition~(\ref{trace-3-symbols}),
fulfilling of which together with conditions
(\ref{hermicity-symbols})--(\ref{trace-2-symbols})
guarantees the operators $\hat{\Pi}_i$ to be rank-1 projectors.

In the considered quantization scheme, the following condition is
to be valid for all $i=1,\ldots,d^2$:
\begin{eqnarray}
\label{trace-3-symbols-Q} (\ref{trace-3-symbols}) ~
\Longleftrightarrow ~ V_i(\|Q\|) := \sum_{p,q,r=1}^{d^2} \left(
\|\mathfrak{S}\|^{\rm T} \|Q\| \|\mathscr{S}\| \right)_{pi} \left(
\|\mathfrak{S}\|^{\rm T} \|Q\| \|\mathscr{S}\| \right)_{qi} \left(
\|\mathfrak{S}\|^{\rm T} \|Q\| \|\mathscr{S}\| \right)_{ri}
\|\mathscr{D}_{p}\|_{qr} = 1, \nonumber\\
\end{eqnarray}
\noindent where we introduced a $d^2$$\times $$d^2$ matrix
$\|\mathscr{D}_{p}\|$ with known matrix elements
\begin{equation}
\|\mathscr{D}_{p}\|_{qr} = {\rm Tr} \big[ \hat{D}\big( (L,k) = p
\big) \hat{D}\big( (L,k) = q \big) \hat{D}\big( (L,k) = r \big)
\big].
\end{equation}

Indeed, this is the restriction (\ref{trace-3-symbols-Q}) that
specifies the orthogonal matrix $\|Q\|$ and, consequently, the
SIC projectors $\hat{\Pi}_i$. Since 1 is a maximum possible value
of the functional $V_i(\|Q\|)$ for any $i=1,\ldots,d^2$, we can
now introduce an operational definition of the desired orthogonal
matrix $\|Q_{\rm SIC}\|$
\begin{equation}
\label{Q-maximum} \max_{\|Q\| \in (\ref{Q-block})} \left\{
\sum_{i=1}^{d^2} V_i(\|Q\|) \right\} = \sum_{i=1}^{d^2} V_i(
\|Q_{\rm SIC}\| ) = d^2.
\end{equation}

Alternatively, one can determine $\|Q_{\rm SIC}\|$ as a solution
of the nonlinear matrix-like equation
\begin{equation}
\label{Q-matrix-equation} \Big( \| \mathscr{S} \|^{\rm T} \| Q
\|^{\rm T} \|\mathfrak{S}\| \|\mathscr{D}_{p}\|
\|\mathfrak{S}\|^{\rm T} \| Q \| \| \mathscr{S} \| \Big)_{ii} =
\Big( \| \mathscr{S} \|^{\rm T} \| Q \|^{\rm T} \|\mathfrak{S}\|
\Big)_{ip},
\end{equation}
\noindent which is nothing else but a reflection of the fact that
$\hat{\Pi}_i^2 = \hat{\Pi}_i$ for all $i=1,\ldots,d^2$.

Finally, the orthogonal matrix $\|Q_{\rm SIC}\|$ is given by a
$(d^2-1)\times(d^2-1)$ block $\|\widetilde{Q}_{\rm SIC}\|$ in
formula~(\ref{Q-block}) which, in turn, can be represented in the
form of sequential rotations of Euclidean space
$\mathbb{R}^{d^2-1}$ (discussed also in the subsequent
section). In fact, $\|\widetilde{Q}_{\rm SIC}\| =
\|\widetilde{Q}_{d^2-1}\| \cdots \|\widetilde{Q}_{2}\|
\|\widetilde{Q}_{1}\|$, where $\|\widetilde{Q}_{1}\|$ is chosen in
such a way that operator $\hat{\Pi}_1$ becomes
positive-semidefinite [and, consequently, a rank-1 projector in
view of already fulfilled requirements~(\ref{hermicity-symbols})--(\ref{trace-2-symbols})].
Then, the rotation $\|\widetilde{Q}_{2}\|$ is applied that remains the
projector $\hat{\Pi}_1$ undisturbed. The rotation angle is chosen
for the operator $\hat{\Pi}_2$ to be nonnegative, and so on. At
the $i$th step, the rotation $\|\widetilde{Q}_{i}\|$ changes
operators $\{\hat{\Pi}_k\}_{k=i}^{d^2}$ only, determines the
explicit form of a new projector $\hat{\Pi}_i$, and leaves all the
found projectors $\{\hat{\Pi}_k\}_{k=1}^{i-1}$ the same.

\subsection{\label{subsection-equiangular} Equiangular Vectors in Euclidean Space}
\begin{figure}
\includegraphics{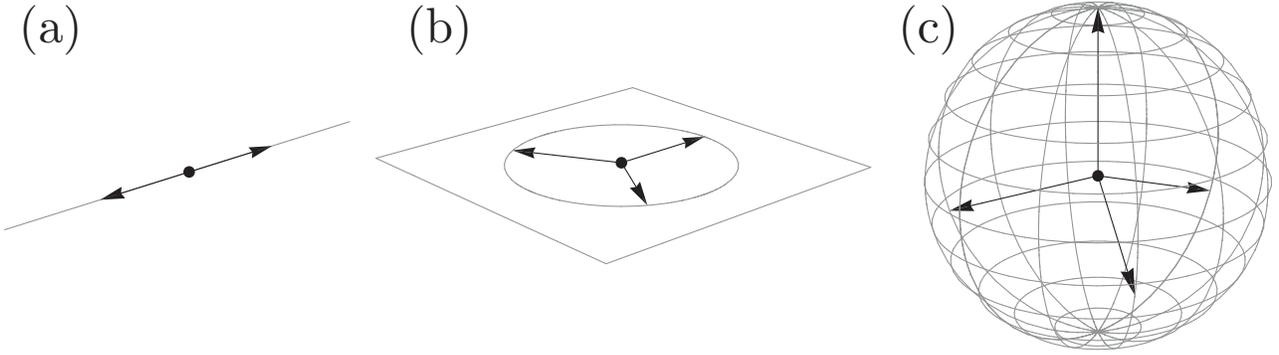}
\caption{\label{figure1} Equiangular vectors in Euclidean spaces:
one-dimensional $\mathbb{R}^1$~(a), two-dimensional $\mathbb{R}^2$~(b),
and three-dimensional $\mathbb{R}^3$~(c). If the dimension
$N$ of space $\mathbb{R}^{N}$ satisfies the condition $N+1=d^2$
for some integer $d$, then such equiangular vectors illustrate
the SIC projectors $\{\hat{\Pi}_i\}_{i=1}^{d^2}$ in the Hilbert space
$\mathscr{H}_d$ ($\mathbb{R}^{d^2-1} \sim \mathscr{H}_d$).}
\end{figure}
\pst It is worth mentioning that we have solved incidentally the
problem of finding equiangular vectors in Euclidean space
$\mathbb{R}^{N}$ (the problem is formulated in wide sense in
\cite{lemmens}). Indeed, according to general formula~(\ref{S-scr-general}),
one can introduce the square matrix $\|\mathscr{S}_{N+1}\|$ of any
dimension $(N+1) \times (N+1)$ by assuming $N+1=d^2$
(here, $d$ is not necessarily an integer). Such
a matrix can be rewritten as
\begin{equation}
\|\mathscr{S}_{N+1}\| = \left(%
\begin{array}{ccccc}
  \displaystyle{\frac{1}{\sqrt[4]{N+1}}} &
  \displaystyle{\frac{1}{\sqrt[4]{N+1}}} &
  \displaystyle{\frac{1}{\sqrt[4]{N+1}}} &
  \cdots & \displaystyle{\frac{1}{\sqrt[4]{N+1}}} \\[3mm]
\hline \\[-5mm]
  {\bf r}_{1}^{(N)} & {\bf r}_{2}^{(N)} & {\bf r}_{3}^{(N)} & \cdots & {\bf r}_{N+1}^{(N)} \\
\end{array}%
\right),
\end{equation}
\noindent where we introduced $N$-dimensional vectors ${\bf
r}_i^{(N)} \in \mathbb{R}^{N}$, $i=1,\ldots,N+1$ such that the
scalar product $({\bf r}_i^{(N)} \cdot {\bf r}_j^{(N)}) = {\rm
const}$ for all $i \ne j$. The explicit form of vectors $\{{\bf
r}_i^{(N)}\}_{i=1}^{N+1}$ is readily achieved from
formula~(\ref{S-scr-general}) by replacing $d \rightarrow
\sqrt{N+1}$. Meanwhile, a similar construction \cite{tremain} came
to our attention. The vectors $\{{\bf r}_i^{(N)}\}_{i=1}^{N+1}$
are illustrated in Euclidean spaces $\mathbb{R}^1$,
$\mathbb{R}^2$, and $\mathbb{R}^3$ in Fig.~\ref{figure1}. The
latter case implies $N+1=4=2^2$, so it gives a simple illustration
of the SIC projectors inside the Bloch ball in $\mathscr{H}_2$.
Continuing this line of reasoning, we see that SIC projectors can
be written in the form $\hat{\Pi}_i = \frac{1}{d} \hat{I}_{d\times
d} + \left(\hat{\bf E} \cdot {\bf r}_i^{(d^2-1)} \right)$, where
$\hat{\bf E}$ is a vector operator composed of $(d^2-1)$ traceless
mutually orthogonal operators $\hat{E}_k$, ${\rm Tr} \big[
\hat{E}_k^2 \big] =1$, defined through generally unknown
orthogonal matrix $\|Q_{\rm SIC}\|$ as follows:
\begin{equation}
\left(%
\begin{array}{c}
  \frac{1}{\sqrt{d}}\hat{I} \\
  \hat{\bf E} \\
\end{array}%
\right) =  \| Q_{\rm SIC} \|^{\rm T} \|\mathfrak{S}\| \hat{\bf D}
= \| Q_{\rm SIC} \|^{\rm T} \|\mathfrak{S}^{-1}\|^{\rm T} \hat{\bf
S}.
\end{equation}

Let us now recall, that unitarily equivalent SIC-POVMs are
obtained from each other by applying a unitary transformation
$\hat{u}$, i.e., $\hat{\Pi}_i \rightarrow
\hat{u}\hat{\Pi}_i\hat{u}^{\dag}$. In the picture developed, such
a transformation results in the transition from one orthogonal
basis of operators $\hat{\bf E}$ to the other, i.e., $\hat{E}_i
\rightarrow \hat{u}\hat{E}_i\hat{u}^{\dag}$. As far as orthogonal
matrix $\|Q_{\rm SIC}\|$ is concerned, such a transformation is
equivalent to multiplication $\|Q_{\rm SIC}\| \rightarrow
\|Q_{u}\| \|Q_{\rm SIC}\|$ by an orthogonal matrix $\|Q_u\|$  with
matrix elements $\|Q_u\|_{kl}= \sum_{p,q=1}^{d^2}
\|\mathfrak{S}^{-1}\|_{pl} \|\mathfrak{S}^{-1}\|_{qk} {\rm Tr}
\big[ \hat{u} \hat{S}_{p} \hat{u}^{\dag} \hat{S}_{q} \big]$. Thus,
unitary transformation $\hat{u}$ rotates vectors ${\bf
r}_i^{(d^2-1)}$ in the Euclidean space $\mathbb{R}^{d^2-1}$ in
accordance with the rule ${\bf r}_i^{(d^2-1)} \rightarrow
\sum_{k=1}^{d^2} \big( \|Q_{\rm SIC}\|^{\rm T} \|Q_{u}\| \|Q_{\rm
SIC}\| \big)_{ik} {\bf r}_k^{(d^2-1)}$.

The last remark of this section is that the classification of
SIC-POVMs is closely related to the properties of the orthogonal
matrix $\|Q_{\rm SIC}\|$.

\section{\label{section-sic-star-product} SIC Star-Product Quantization Scheme}
\pst In this section, we consider the SIC star-product
quantization scheme outlined concisely in
Sec.~\ref{subsection-SIC-de-quantizer}. The main idea is that,
although the exact form of SIC projectors $\hat{\Pi}_i$ is not
known in general dimension, it is still possible to derive some
extra properties of projectors $\hat{\Pi}_i$ by assuming that
operators
\begin{equation}
\hat{U}_i = \frac{1}{d}\hat{\Pi}_i {\rm \qquad and \qquad}
\hat{D}_i = (d+1)\hat{\Pi}_i - \hat{I} , \qquad i=1,\ldots, d^2
\end{equation}
\noindent are indeed the dequantizer and quantizer, respectively
(i.e., define a true quantization scheme). For instance, let us
consider properties (\ref{kernel-3}), (\ref{kernel-4}) of the star-product
kernel $K_{ijk} = {\rm Tr} \big[ \hat{D}_i \hat{D}_j\hat{U}_k \big]$
and express them in terms of SIC-projectors' triple product
$T_{ijk} = {\rm Tr} \big[ \hat{\Pi}_i \hat{\Pi}_j\hat{\Pi}_k \big]$.

To start, using relation~(\ref{scalarproduct}) it is not hard
to see that the delta-function [in view of formula~(\ref{star-product-delta})]
on SIC tomographic symbols reads
$\mathfrak{D}_{ij}={\rm Tr} \big[ \hat{U}_i \hat{D}_j \big] =
\delta_{ij}$, i.e., reduces to a conventional Kronecker delta-symbol.
Further, the star-product kernel reads
\begin{equation}
\label{K-through-T} K_{ijk} = \frac{1}{d} \left[ (d+1)^2 T_{ijk}
-d(\delta_{ik}+\delta_{jk}) -1 \right].
\end{equation}
\noindent Now, it is possible to calculate the higher-order
star-product kernels $K_{ijkl}^{(3)}$ and $K_{ijklm}^{(4)}$ by
virtue of several different expressions [see
(\ref{kernel-3}), (\ref{kernel-4})]. We omit the intermediate
calculations and present, as a result of this consideration, the
necessary condition for $T_{ijk}$ to obey
\begin{eqnarray}
&& \label{T-3-kernel}\sum_{m=1}^{d^2} \left( T_{ijm}T_{mkl} -
T_{iml}T_{jkm} \right) = \frac{d}{(d+1)^3} \Big[
(d\delta_{ij}+1)(d\delta_{kl}+1) -
(d\delta_{jk}+1)(d\delta_{il}+1) \Big],\\
&& \label{T-4-kernel-Eq1} \sum_{n,p=1}^{d^2} \left(
T_{ijn}T_{nkp}T_{plm} - T_{inp}T_{jkn}T_{plm} \right) =
\frac{(d+1)^2}{d^2} \Bigg[ \frac{d\delta_{ij}+1}{d+1} T_{klm} +
\frac{(d\delta_{ij}+1)(d\delta_{lm}+1)}{(d+1)^2} \nonumber\\
&&
\qquad\qquad\qquad\qquad\qquad\qquad\qquad\qquad\qquad\qquad\qquad
- T_{ilm}\frac{d\delta_{jk}+1}{d+1} -
\frac{(d\delta_{jk}+1)(d\delta_{lm}+1)}{(d+1)^2} \Bigg],\\
&& \label{T-4-kernel-Eq2} \sum_{n,p=1}^{d^2} \left(
T_{ijn}T_{nkp}T_{plm} - T_{ijp}T_{kln}T_{pnm} \right) =
\frac{(d+1)^2}{d^2} \Bigg[ T_{ijk}\frac{d\delta_{lm}+1}{d+1} +
\frac{(d\delta_{ij}+1)(d\delta_{lm}+1)}{(d+1)^2} \nonumber\\
&&
\qquad\qquad\qquad\qquad\qquad\qquad\qquad\qquad\qquad\qquad\qquad
- T_{ijm}\frac{d\delta_{kl}+1}{d+1} -
\frac{(d\delta_{ij}+1)(d\delta_{kl}+1)}{(d+1)^2} \Bigg],
\qquad\qquad
\end{eqnarray}
\noindent where formula (\ref{T-3-kernel}) corresponds to
equality~(\ref{kernel-3}), formulas~(\ref{T-4-kernel-Eq1}) and
(\ref{T-4-kernel-Eq2}) are responsible for two of five
equalities~(\ref{kernel-4}), with the other being re-expressed
through the presented ones. Also, using two expressions for a
star-product kernel, namely, the definition
(\ref{star-product-kernel}) and expansions (\ref{kernel-3}) and
(\ref{kernel-4}), we succeeded in establishing a relation between
the triple-product $T_{ijk}$ and the higher-order products, e.g.,
the four-product ${\rm Tr}\big[ \hat{\Pi}_i \hat{\Pi}_j
\hat{\Pi}_k \hat{\Pi}_l \big]$ and the five-product ${\rm Tr}\big[
\hat{\Pi}_i \hat{\Pi}_j \hat{\Pi}_k \hat{\Pi}_l \hat{\Pi}_m
\big]$. The result is
\begin{eqnarray}
&& {\rm Tr}\big[ \hat{\Pi}_i \hat{\Pi}_j \hat{\Pi}_k \hat{\Pi}_l
\big] = \frac{d+1}{d} \sum_{m=1}^{d^2} T_{ijm}T_{mkl} -
\frac{(d\delta_{ij}+1)(d\delta_{kl}+1)}{(d+1)^2} , \\
&& {\rm Tr}\big[ \hat{\Pi}_i \hat{\Pi}_j \hat{\Pi}_k \hat{\Pi}_l
\hat{\Pi}_m \big] =  \frac{(d+1)^2}{d^2} \sum_{n,p=1}^{d^2}
T_{ijn}T_{nkp}T_{plm} \nonumber\\
&& \label{five-product}\qquad\qquad\qquad\qquad\qquad\qquad\qquad
- T_{ijk}\frac{d\delta_{lm}+1}{d+1} -
\frac{(d\delta_{ij}+1)(d\delta_{lm}+1)}{(d+1)^2} -
\frac{d\delta_{ij}+1}{d+1}T_{klm}.\qquad\qquad
\end{eqnarray}

It is also worth mentioning that the same relations can be
obtained (even in an easier way) with the help of the star-product
kernel for dual symbols~(\ref{star-product-kernel-dual}) which, in
our case, is
\begin{equation}
\label{K-dual-through-T} K_{ijk}^{\rm dual} = \frac{1}{d^2(d+1)}
\left[ (d+1)^2 T_{ijk} - (d\delta_{ij} + 1) \right].
\end{equation}

\section{\label{section-Qubit} SIC-POVMs for Qubits}
\pst Let us now develop the above approach to SIC-POVMs for qubits
($d=2$).

To anticipate, the SIC projectors can be chosen as follows:
\begin{eqnarray}
\label{sic-proj-qubit-apriori}
&& \hat{\Pi}_1 = \frac{1}{2\sqrt{3}} \left(%
\begin{array}{cc}
  \sqrt{3}+1 & 1-i \\
  1+i & \sqrt{3}-1 \\
\end{array}%
\right), \qquad \hat{\Pi}_2 = \frac{1}{2\sqrt{3}} \left(%
\begin{array}{cc}
  \sqrt{3}-1 & 1+i \\
  1-i & \sqrt{3}+1 \\
\end{array}%
\right),\nonumber\\[-2mm]
&&\\[-2mm]
&& \hat{\Pi}_3 = \frac{1}{2\sqrt{3}} \left(%
\begin{array}{cc}
  \sqrt{3}-1 & -1-i \\
  -1+i & \sqrt{3}+1 \\
\end{array}%
\right), \qquad \hat{\Pi}_4 = \frac{1}{2\sqrt{3}} \left(%
\begin{array}{cc}
  \sqrt{3}+1 & -1+i \\
  -1-i & \sqrt{3}-1 \\
\end{array}%
\right).\nonumber
\end{eqnarray}
\noindent It is readily seen that ${\rm Tr}\big[ \hat{\Pi}_i \big]
={\rm Tr}\big[ \hat{\Pi}_i^2 \big] = {\rm Tr}\big[ \hat{\Pi}_i^3
\big] = 1$ for all $i=1,\ldots,4$, and ${\rm Tr}\big[ \hat{\Pi}_i
\hat{\Pi}_j \big] = 1/3$ if $i \ne j$. The former equality
guaranties $\hat{\Pi}_i$ to be rank-1 projectors, whereas the
latter one shows they are symmetric. The corresponding equiangular
vectors in $\mathbb{C}^2$ read
\begin{eqnarray}
&& |\psi_1\rangle =  \frac{1}{\sqrt{{2\sqrt{3}}}}\left(%
\begin{array}{l}
  \sqrt{\sqrt{3}+1} \\
  \sqrt{\sqrt{3}-1} ~ e^{\mathfrak{i}\pi/4} \\
\end{array}%
\right), \qquad |\psi_2\rangle = \frac{1}{\sqrt{{2\sqrt{3}}}} \left(%
\begin{array}{l}
  \sqrt{\sqrt{3}-1} \\
  \sqrt{\sqrt{3}+1} ~ e^{-\mathfrak{i}\pi/4} \\
\end{array}%
\right), \nonumber\\[-2mm]
&&\\[-2mm]
&& |\psi_3\rangle = \frac{1}{\sqrt{{2\sqrt{3}}}} \left(%
\begin{array}{l}
  \sqrt{\sqrt{3}-1} \\
  \sqrt{\sqrt{3}+1} ~ e^{\mathfrak{i} 3\pi/4} \\
\end{array}%
\right), \qquad |\psi_4\rangle = \frac{1}{\sqrt{{2\sqrt{3}}}} \left(%
\begin{array}{l}
  \sqrt{\sqrt{3}+1} \\
  \sqrt{\sqrt{3}-1} ~ e^{-\mathfrak{i}3\pi/4} \\
\end{array}%
\right).\nonumber
\end{eqnarray}

\subsection{Seeking SIC-POVMs in Dimension $\boldsymbol{d=2}$}
\pst To start, we find all possible SIC-POVMs by employing formula
(\ref{Pi-vector-found}). The explicit form of the 4$\times$4
matrix $\|\mathscr{S}\|$ reads
\begin{equation}
\label{S-scr-qubit}
\|\mathscr{S}\| = \left(%
\begin{array}{cccc}
  \frac{1}{\sqrt{2}} & \frac{1}{\sqrt{2}} & \frac{1}{\sqrt{2}} & \frac{1}{\sqrt{2}} \\
  -\frac{1}{\sqrt{3}} & \frac{1}{\sqrt{3}} & 0 & 0 \\
  -\frac{1}{3} & -\frac{1}{3} & \frac{2}{3} & 0 \\
  -\frac{1}{3\sqrt{2}} & -\frac{1}{3\sqrt{2}} & -\frac{1}{3\sqrt{2}} & \frac{1}{\sqrt{2}} \\
\end{array}%
\right).
\end{equation}
\noindent The $d^2$$\times$$d^2$ matrix $\|\mathfrak{S}\|$ is given
by formulas (\ref{S-frak})--(\ref{Sigma-blocks}), which at
$d=2$ yield
\begin{equation}
\label{S-frak-qubit}
\|\mathfrak{S}\| = \left(%
\begin{array}{cccc}
  1 & 0 & 0 & 0 \\
  0 & \cos\theta_1 & \cos\theta_2 & \cos\theta_3 \\
  0 & \cos\varphi_1\sin\theta_1 & \cos\varphi_2\sin\theta_2 & \cos\varphi_3\sin\theta_3 \\
  0 & \sin\varphi_1\sin\theta_1 & \sin\varphi_2\sin\theta_2 & \sin\varphi_3\sin\theta_3 \\
\end{array}%
\right) = \left(%
\begin{array}{cccc}
  1 & 0 & 0 & 0 \\
  0 & n_{1z} & n_{2z} & n_{3z} \\
  0 & n_{1x} & n_{2x} & n_{3x} \\
  0 & n_{1y} & n_{2y} & n_{3y} \\
\end{array}%
\right), \qquad\quad
\end{equation}
\noindent whereas, according to formula
(\ref{discrete-Chebyshev}), the vector operator $\hat{\bf S}$
reads
\begin{equation}
\label{S-vect-oper-qubit}
\hat{\bf S} \equiv \left(%
\begin{array}{c}
  \hat{S}_0^{(1/2)}({\bf n}_1) \\
  \hat{S}_1^{(1/2)}({\bf n}_1) \\
  \hat{S}_1^{(1/2)}({\bf n}_2) \\
  \hat{S}_1^{(1/2)}({\bf n}_3) \\
\end{array}%
\right) = \left(%
\begin{array}{c}
  \frac{1}{\sqrt{2}}\hat{I} \\
  \sqrt{2}(\hat{\bf J}\cdot {\bf n}_1) \\
  \sqrt{2}(\hat{\bf J}\cdot {\bf n}_2) \\
  \sqrt{2}(\hat{\bf J}\cdot {\bf n}_3) \\
\end{array}%
\right) = \frac{1}{\sqrt{2}} \left(%
\begin{array}{c}
  \hat{I} \\
  (\hat{\boldsymbol\sigma}\cdot {\bf n}_1) \\
  (\hat{\boldsymbol\sigma}\cdot {\bf n}_2) \\
  (\hat{\boldsymbol\sigma}\cdot {\bf n}_3) \\
\end{array}%
\right).
\end{equation}
\noindent Substituting (\ref{S-scr-qubit}) for $\|\mathscr{S}\|$,
(\ref{S-frak-qubit}) for $\|\mathfrak{S}\|$, and
(\ref{S-vect-oper-qubit}) for $\hat{\bf S}$ in
(\ref{Pi-vector-found}), after simplification, we obtain
\begin{equation}
\label{Pi-qubit}
\hat{\boldsymbol\Pi} = \frac{1}{2} \left(%
\begin{array}{cccc}
  1 & -\sqrt{\frac{2}{3}} & -\frac{\sqrt{2}}{3} & -\frac{1}{3} \\
  1 & \sqrt{\frac{2}{3}} & -\frac{\sqrt{2}}{3} & -\frac{1}{3} \\
  1 & 0 & \frac{2\sqrt{2}}{3} & -\frac{1}{3} \\
  1 & 0 & 0 & 1 \\
\end{array}%
\right) \left(%
\begin{array}{c|ccc}
  1 & 0 & 0 & 0\\
  \hline
  0 & \multicolumn{3}{c}{} \\
  0 & \multicolumn{3}{c}{\widetilde{Q}^{\rm T}}\\
  0 & \multicolumn{3}{c}{}\\
\end{array}%
\right) \left(%
\begin{array}{c}
  \hat{I} \\
  \hat{\sigma}_z \\
  \hat{\sigma}_x \\
  \hat{\sigma}_y \\
\end{array}%
\right) = \frac{1}{2} \left(%
\begin{array}{c}
  \hat{I} + (\hat{\boldsymbol\sigma}\cdot \|R\|{\bf r}_1)  \\
  \hat{I} + (\hat{\boldsymbol\sigma}\cdot \|R\| {\bf r}_2) \\
  \hat{I} + (\hat{\boldsymbol\sigma}\cdot \|R\| {\bf r}_3) \\
  \hat{I} + (\hat{\boldsymbol\sigma}\cdot \|R\| {\bf r}_4) \\
\end{array}%
\right), \qquad
\end{equation}
\noindent where we introduce a 3$\times$3 orthogonal matrix
$\|R\|$ and normalized vectors $\{{\bf r}_i\}_{i=1}^{4} \in
\mathbb{R}^3$ given by
\begin{equation}
\label{vectors-qubit}
\|R\| = \left(%
\begin{array}{ccc}
  0 & 1 & 0 \\
  0 & 0 & 1 \\
  1 & 0 & 0 \\
\end{array}%
\right)\|\widetilde{Q}\|, ~ {\bf r}_1 = \left(%
\begin{array}{c}
  -\sqrt{\frac{2}{3}} \\
  -\frac{\sqrt{2}}{3} \\
  -\frac{1}{3} \\
\end{array}%
\right), ~ {\bf r}_2 = \left(%
\begin{array}{c}
  \sqrt{\frac{2}{3}} \\
  -\frac{\sqrt{2}}{3} \\
  -\frac{1}{3} \\
\end{array}%
\right), ~{\bf r}_3 = \left(%
\begin{array}{c}
0 \\
\frac{2\sqrt{2}}{3} \\
-\frac{1}{3} \\
\end{array}%
\right), ~ {\bf r}_4 = \left(%
\begin{array}{c}
0 \\
0 \\
1 \\
\end{array}%
\right). \quad
\end{equation}

It is shown in Sec.~\ref{subsection-search} that, if the set of
operators $\{\hat{\Pi}_i\}_{i=1}^{4}$ is constructed in such a
way, then the hermicity condition~(\ref{hermicity-symbols}) as
well as the trace conditions~(\ref{trace-1-symbols}) and
(\ref{trace-2-symbols}) of the first and second orders,
respectively, is fulfilled automatically. For
$\{\hat{\Pi}_i\}_{i=1}^{4}$ to be true SIC projectors, the extra
restriction on orthogonal matrix $\|Q\|$ is imposed by
requirement~(\ref{Q-maximum}) or an equivalent
requirement~(\ref{Q-matrix-equation}). In our case, these
restrictions can be easily written in terms of the 3$\times$3
matrix $\|R\|$. Surprisingly enough that, in the case of qubits,
these additional conditions are also fulfilled for an
\textit{arbitrary} orthogonal matrix $\|R\|$. This means that
formula~(\ref{Pi-qubit}) gives the explicit solution to the
problem of SIC existence, with one SIC-POVM construction differing
from the other in the choice of the orthogonal matrix $\|R\|$
only. The geometrical sense of this fact in the Bloch ball picture
in Fig.~\ref{figure1}c is that all possible sets of SIC projectors
$\{\hat{\Pi}_i\}_{i=1}^{4}$ (unambiguously defined by vectors
$\{{\bf r}_i\}_{i=1}^{4}$) are obtained by applying an orthogonal
transformation to all the vectors $\{{\bf r}_i\}_{i=1}^{4}$
simultaneously. For example, the SIC
projectors~(\ref{sic-proj-qubit-apriori}) are obtained from
formula~(\ref{Pi-qubit}) by an orthogonal transformation $\|R\|$
which transforms vectors~(\ref{vectors-qubit}) into vectors
$\frac{1}{\sqrt{3}}(1,1,1)$, $\frac{1}{\sqrt{3}}(1,-1,-1)$,
$\frac{1}{\sqrt{3}}(-1,1,-1)$, and $\frac{1}{\sqrt{3}}(-1,-1,1)$.
Among all possible orthogonal transformations $\|R\|$, one can
point out the case of rotation ($\det \|R\|=1$) and inversion
($\det \|R\|=-1$). Therefore, we have two SIC-POVM sets generated
by fixed representation operator, namely, the Weyl--Heisenberg
displacement operator. To be concise, we have two different
fiducial vectors, such that one vector can be obtained from the
other by inversion of the Bloch ball picture~\cite{renes}.

\subsection{\label{subsection-qubit-kernel} Star-Product Kernel for Qubits}
\pst Now, when the SIC projectors are found, let us explore their
properties.

To start, we focus our attention on the star-product kernel
$K_{ijk}$ which is connected with many additional relations, e.g.,
(\ref{T-3-kernel})--(\ref{five-product}). Employing well-known
properties of Pauli matrices, it is not hard to see that the
triple-product $T_{ijk}={\rm Tr}\big[ \hat{\Pi}_i \hat{\Pi}_j
\hat{\Pi}_k \big]$ reads
\begin{equation}
T_{ijk} = \frac{1}{4} \Big\{ 1 + ({\bf r}_i \cdot {\bf r}_j) +
({\bf r}_j \cdot {\bf r}_k) + ({\bf r}_k \cdot {\bf r}_i) +
\mathfrak{i} ({\bf r}_i \cdot [{\bf r}_j \times {\bf r}_k])\Big\},
\end{equation}
\noindent where $\mathfrak{i}$ is the imaginary unit and $({\bf
r}_i \cdot [{\bf r}_j \times {\bf r}_k])$ denotes the standard
triple product of vectors. Further, we take into account that, in
our case, $({\bf r}_i \cdot {\bf r}_j) = (4\delta_{ij}-1)/3$ and
$({\bf r}_i \cdot [{\bf r}_j \times {\bf r}_k]) =
-4\varepsilon_{ijk}/3\sqrt{3}$, where $\varepsilon_{ijk}$ is an
antisymmetric tensor such that
$\varepsilon_{123}=\varepsilon_{134}=\varepsilon_{142}=\varepsilon_{432}=1$.
Finally, using (\ref{K-through-T}) and (\ref{K-dual-through-T}),
we obtain
\begin{eqnarray}
&& T_{ijk} = \frac{1}{3} \Big\{
\delta_{ij}+\delta_{jk}+\delta_{ki} -\frac{\mathfrak{i}}{\sqrt{3}}
\varepsilon_{ijk} \Big\}, \qquad K_{ijk} = \frac{1}{2} \Big\{ 3
\delta_{ij}-\mathfrak{i}\sqrt{3} \varepsilon_{ijk} -1 \Big\}, \\
&& K_{ijk}^{\rm dual} = \frac{1}{12} \Big\{ \delta_{ij} +
3(\delta_{jk}+\delta_{ki}) -\mathfrak{i}\sqrt{3} \varepsilon_{ijk}
-1 \Big\}.
\end{eqnarray}
\noindent It is known that there exists a recurrence relation on
star-product kernels of spin-tomographic symbols $w(m,{\bf n})$
(see the family of spin-tomographic symbols in
Sec.~\ref{section-generic-star-product-scheme}) which connects the
kernel of spin $j$ with the kernels of spins $(j-1/2)$ and
$(j-1)$~\cite{filipp-chebyshev}. We hope that a similar relation
does exist for SIC tomographic kernels $K_{ijk}$ as well. In this
case, such a relation would connect kernels for dimensions $d$,
$(d-1)$, and $(d-2)$.

\subsection{\label{subsection-intertwining}Intertwining Kernels to Other Quantization Schemes}
\pst We emphasize here that the SIC representation of quantum
states (Sec.~\ref{subsection-SIC-representation}) is closely
related to other probability representations of quantum mechanics.
This means that there exists a relation between the SIC
quantization scheme and other quantization schemes. This section
is devoted to establishing such a relation between SIC tomographic
symbols $f_A(i)\equiv {\rm Tr} \big[ \hat{A} \hat{U}_i \big]$ and
spin-tomographic symbols $f_A(m,{\bf n})$ as well as spin-FNR
tomographic symbols $f_A(m,{\bf n}_k)$ (spin tomography with a
finite number of rotations) outlined briefly in
Sec.~\ref{section-generic-star-product-scheme}.

In fact, utilizing general relations~(\ref{symbol}) and
(\ref{A-from-symbol}) of the star product, we immediately obtain
the following relation between two quantization schemes ${\bf
x}_1$ and ${\bf x}_2$:
\begin{equation}
\label{intertwining-general} f_A({\bf x}_2) = \int f_A({\bf x}_1)
{\rm Tr}\big[ \hat{D}({\bf x}_1) \hat{U}({\bf x}_2) \big] d{\bf
x_1}.
\end{equation}
\noindent We will refer to the kernel $\mathcal{K}_{1 \rightarrow
2}({\bf x}_1,{\bf x}_2) := {\rm Tr}\big[ \hat{D}({\bf x}_1)
\hat{U}({\bf x}_2) \big]$ as the intertwining kernel between
schemes ${\bf x}_1$ and ${\bf x}_2$. Applying formula~(\ref{intertwining-general})
to qubits, we readily obtain the explicit form of the intertwining
kernels between the SIC quantization scheme [determined by
SIC projectors~(\ref{Pi-qubit})] and alternative quantization schemes
given by (\ref{dequant-orth}), (\ref{dequant-FNR}), (\ref{quant-FNR}),
and vice versa. The result is
\begin{eqnarray}
&& \mathcal{K}_{\rm Spin \rightarrow  SIC}(m,{\bf n}, i) =
\frac{1}{4} \Big( 1+6m({\bf n} \cdot \|R\| {\bf r}_i) \Big),\\
&& \mathcal{K}_{\rm SIC \rightarrow Spin}(i,m,{\bf n}) =
\frac{1}{2} \Big( 1+6m({\bf n} \cdot \|R\| {\bf r}_i) \Big),\\
&& \label{Spin-FNR-SIC}\mathcal{K}_{\rm SpinFNR \rightarrow
SIC}(m,{\bf n}_k, i) = \frac{3}{4} \Big( \delta_{k,1} + 2m({\bf
l}_k \cdot \|R\| {\bf r}_i)
\Big),\\
&& \mathcal{K}_{\rm SIC \rightarrow SpinFNR}(i,m,{\bf n}_k) =
\frac{1}{6} \Big( 1 + 6m({\bf n}_k \cdot \|R\| {\bf r}_i) \Big),
\end{eqnarray}
\noindent with vectors ${\bf l}_k$, $k=1,2,3$ in Eq.~(\ref{Spin-FNR-SIC})
forming a dual basis with respect to the directions
$\{{\bf n}_k\}_{k=1}^{3}$~\cite{serg-inverse-spin}.

\subsection{\label{subsection-MUB} Relation to Mutually Unbiased Bases}
\pst It is known that the SIC construction and so-called mutually
unbiased bases (MUBs) have some common
properties~\cite{albouy,appleby-MUBs}. In order to illustrate this
relation, we restrict ourselves to the case of qubits only. Using
the standard notation $|\uparrow\rangle:=|j=1/2,m=1/2\rangle$,
$|\downarrow\rangle:=|j=1/2,m=-1/2\rangle$, one can introduce
three bases $\big\{ |\mu_i\rangle, |\nu_i\rangle \big\}_{i=1}^{3}$
such that $\langle\mu_i | \nu_i\rangle =0$ for all $i=1,2,3$ and
$\left| \langle\mu_i | \mu_j\rangle \right|^2 = \left|
\langle\mu_i | \nu_j\rangle \right|^2 = \left| \langle\nu_i |
\nu_j\rangle \right|^2 = 1/2$ for all $i \ne j$. Indeed, a
possible choice of MUBs is as follows:
\begin{eqnarray}
\label{MUBs} && |\mu_1\rangle = |\uparrow\rangle, \quad
|\nu_1\rangle = |\downarrow\rangle, \qquad |\mu_2\rangle =
\frac{|\uparrow\rangle+|\downarrow\rangle}{\sqrt{2}}, \quad
|\nu_2\rangle
=\frac{|\uparrow\rangle-|\downarrow\rangle}{\sqrt{2}},
\nonumber\\[-2mm]
&&\\[-2mm]
&& |\mu_3\rangle =
\frac{|\uparrow\rangle+\mathfrak{i}|\downarrow\rangle}{\sqrt{2}},
\quad |\nu_3\rangle =
\frac{|\uparrow\rangle-\mathfrak{i}|\downarrow\rangle}{\sqrt{2}}.\nonumber
\end{eqnarray}
The states (\ref{MUBs}) correspond to the following points on the
Bloch sphere: $(0,0,1)$, $(0,0,-1)$, $(1,0,0)$, $(-1,0,0)$,
$(0,1,0)$, and $(0,-1,0)$, respectively (see Fig.~\ref{figure1}c).
These points are nothing else but vertices of the octahedron. Each
basis $\big\{ |\mu_i\rangle, |\nu_i\rangle \big\}$ corresponds to
the $i$th diagonal of the octahedron. In view of this, similar to
SIC-POVMs which are associated with diagonals of the cube, the
MUBs are also associated with equiangular lines, namely, diagonals
of the octahedron. It is worth mentioning that the problem of
existence of MUBs in $\mathscr{H}_d$ of any dimension $d$ can also
be formulated in terms of symbols of operators.

\section{\label{section-Lie}Notes to Lie Algebraic Consideration}
\pst It is emphasized in \cite{fuchs-2010} that the
$d^2$ operators $\{\hat{\Pi}_i\}_{i=1}^{d^2}$ form a basis for the
complex Lie algebra ${\rm gl}(d,\mathbb{C})$ and satisfy the
commutation relation of the form
\begin{equation}
\label{commutator-Pi} \big[ \hat{\Pi}_i,\hat{\Pi}_j \big] =
\sum_{k=1}^{d^2} J_{ijk} \hat{\Pi}_k.
\end{equation}
\noindent Structure constants $J_{ijk}$ are expressed through the
kernel $K_{ijk}$ as follows:
\begin{equation}
\label{J-through-K} J_{ijk} = \frac{1}{d+1} \left( K_{ijk} -
K_{jik} \right).
\end{equation}
\noindent Formula~(\ref{J-through-K}) is a standard relation
between the structure constants of associative product and the structure
constants of the Lie product.  In order to derive new relations on
projectors $\hat{\Pi}_i$ originating from their Lie algebraic
structure, we will consider case $d=2$ in detail. In this case, we
have connection with the Lie group ${\rm GL}(2,\mathbb{C})$. This
group can be considered as a direct product of the group ${\rm
SL}(2,\mathbb{C})$ and the group of complex numbers with the standard
multiplication rule. The matrices that belong to the group ${\rm
SL}(2,\mathbb{C})$ are the matrices of the Lorentz-group
representation. Generators $H_1$, $H_2$, $H_3$, $F_1$, $F_2$, and
$F_3$ of ${\rm SL}(2,\mathbb{C})$ read~\cite{gelfand-minlos-shapiro}
\begin{equation}
H_1=\mathfrak{i}F_1=\frac{1}{2}\left(%
\begin{array}{cc}
  0 & 1 \\
  1 & 0 \\
\end{array}%
\right), \quad H_2=\mathfrak{i}F_2=\frac{1}{2}\left(%
\begin{array}{cc}
  0 & \mathfrak{i} \\
  -\mathfrak{i} & 0 \\
\end{array}%
\right), \quad H_3=\mathfrak{i}F_3=\frac{1}{2}\left(%
\begin{array}{cc}
  1 & 0 \\
  0 & -1 \\
\end{array}%
\right)
\end{equation}
\noindent and satisfy commutation relations of the form
\begin{eqnarray}
&& \big[ H_{\pm},H_3 \big] = \mp H_{\pm},\quad \big[ H_{+},H_{-}
\big] = 2 H_3, \quad \big[ F_{\pm},F_3 \big] = \pm H_{\pm}, \quad
\big[ F_{+},F_{-} \big] = - 2 H_3,\nonumber\\[-2mm]
&&\\[-2mm]
&& \big[ F_{+},H_{+} \big] = \big[ H_{-},F_{-} \big] = \big[
H_{3},F_{3} \big] = 0, \quad \big[ H_{\pm},F_{\mp} \big] = \pm
2F_{3}, \quad \big[ F_{\pm},H_3 \big] = \mp F_{\pm},
\nonumber
\end{eqnarray}
\noindent where $H_{\pm}=H_1 \pm \mathfrak{i}H_2$, $F_{\pm}=F_1
\pm \mathfrak{i}F_2$, and $\mathfrak{i}$ is the imaginary unit.
Further, there exist two Casimir operators $\hat{C}_{1}$ and
$\hat{C}_{2}$ (see the explicit form, e.g., in
\cite{smorodinskii})
\begin{equation}
\label{casimir} \hat{C}_{1} = \hat{\bf H}^2-\hat{\bf F}^2 +
2\mathfrak{i}(\hat{\bf H}\cdot\hat{\bf F}), \qquad \hat{C}_{2} =
\hat{\bf H}^2-\hat{\bf F}^2 - 2\mathfrak{i}(\hat{\bf
H}\cdot\hat{\bf F}),
\end{equation}
\noindent such that $\big[ \hat{C}_{1}, \hat{H}_i \big] = \big[
\hat{C}_{1}, \hat{F}_i \big] = \big[ \hat{C}_{2}, \hat{H}_i \big]
= \big[ \hat{C}_{2}, \hat{F}_i \big] = 0$ for all $i=1,2,3$. It is
worth mentioning that operators $\frac{1}{2}(\hat{C}_1+\hat{C}_2)$
and $\frac{1}{2\mathfrak{i}}(\hat{C}_1-\hat{C}_2)$ have the sense
of Lorentz invariants of the electromagnetic field.

Let us rewrite
these commutation relations in terms of projectors $\hat{\Pi}_i$,
assuming that only the commutation relation~(\ref{commutator-Pi}) is known.
Indeed, we will then obtain a
new restriction onto projectors $\hat{\Pi}_i$ originating from
their Lie algebraic structure.

In the case $d=2$, relation~(\ref{commutator-Pi}) transforms into
$\big[ \hat{\Pi}_i,\hat{\Pi}_j \big] = \pm \sum_{k=1}^{4}
\frac{\mathfrak{i}}{\sqrt{3}}\varepsilon_{ijk} \hat{\Pi}_k$, where
$\varepsilon_{ijk}$ is an antisymmetric tensor such that
$\varepsilon_{123}=\varepsilon_{134}=\varepsilon_{142}=\varepsilon_{432}=1$
and the sign $\pm$ depends on the labeling of the SIC projectors
(we choose the plus sign). Suppose now
\begin{eqnarray}
&& \hat{H}_1 = \mathfrak{i}\hat{F}_1 = \frac{\sqrt{3}}{8} \left(
\hat{\Pi}_1+\hat{\Pi}_2-\hat{\Pi}_3-\hat{\Pi}_4 \right),\nonumber\\
&&\hat{H}_2 = \mathfrak{i}\hat{F}_2 = \frac{\sqrt{3}}{8} \left( -
\hat{\Pi}_1+\hat{\Pi}_2-\hat{\Pi}_3+\hat{\Pi}_4 \right),\\
&& \hat{H}_3 = \mathfrak{i}\hat{F}_3 = \frac{\sqrt{3}}{8} \left(
\hat{\Pi}_1-\hat{\Pi}_2-\hat{\Pi}_3+\hat{\Pi}_4 \right),\nonumber
\end{eqnarray}
\noindent then Casimir operators (\ref{casimir}) take the form
\begin{equation}
\hat{C}_{1} = 0, \qquad \hat{C}_{2} = \frac{3}{16}\left( 3
\sum_{i=1}^{4} \hat{\Pi}_i^2 - \sum_{i \ne j} \hat{\Pi}_i
\hat{\Pi}_j \right).
\end{equation}
\noindent Finally, starting from commutators~(\ref{commutator-Pi})
and using a specific property of Casimir operators, we manage to
obtain the following commutation relations:
\begin{equation}
\label{casimir-final} \left[ \hat{\Pi}_k, 3 \sum_{i=1}^{4}
\hat{\Pi}_i^2 - \sum_{i \ne j} \hat{\Pi}_i \hat{\Pi}_j \right] = 0
{\rm ~for~all~} k=1,\ldots, 4.
\end{equation}
\noindent In fact, relations similar to (\ref{casimir-final}) can
also be derived in higher dimensions. The crucial point is that
the obtained equation is compatible with conditions
(\ref{hermicity})--(\ref{trace}). Thus, if SIC-POVMs do exist, the
whole series of operator conditions
(\ref{hermicity})--(\ref{trace})$\wedge$(\ref{casimir-final},~generalized)
is to be fulfilled simultaneously.

\section{\label{section-conclusions}Conclusions}
\pst To conclude, we present the main results of the paper.

Combining the ideas of a generic star-product scheme with the
SIC-POVM approach to quantum states, we have shown that the
SIC projectors $\hat{\Pi}_i$ can be considered (up to a
normalization factor and the identity operator) as dequantizers
$\hat{U}_i$ and quantizers $\hat{D}_i$ of the SIC star-product
quantization scheme. From this, it follows immediately that
fulfilling of conditions~(\ref{hermicity})--(\ref{trace}) means
the existence of the associative product $K_{ijk}={\rm Tr}\big[
\hat{D}_i \hat{D}_j \hat{U}_k \big]$ which is a solution of
Eqs.~(\ref{kernel-3}) and (\ref{kernel-4}). Moreover,
utilizing the standard equations for a generic star-product
kernel, we have derived some properties of the triple products
$T_{ijk}={\rm Tr} \big[ \hat{\Pi}_i \hat{\Pi}_j \hat{\Pi}_k \big]$
of SIC projectors found in \cite{fuchs-2010}. Thus, we have
interpreted such properties of $T_{ijk}$ as standard properties of
the star-product kernel (including the dual~\cite{oman'ko-vitale}
star-product scheme). From the same point of view, the Lie
algebraic structure found in \cite{fuchs-2010} is an immediate and
known consequence of the antisymmetrized  kernel of associative
product. Further, the problem of SIC-POVM existence is formulated in
terms of symbols of the SIC projectors and the corresponding
kernel of associative product~(\ref{hermicity-symbols})--(\ref{trace-3-symbols}).
The approach to solve the modified problem is also developed. By example
of qubits, we show the similarity between SIC-POVMs and mutually
unbiased bases (MUBs) and hope to clarify this connection
elsewhere.

The other result of this work is the conclusion that the SIC-POVM is
a partial case of the probability representation of quantum states
and can be related to other known kinds of the probability
representations like the spin tomography, unitary tomography,
and spin tomography with a finite number of rotations. Also,
we cannot help mentioning a conceptual drawback of the
SIC representation, namely, the absence of a measurable physical
quantity which can give rise to the SIC probability distribution.

\section*{Acknowledgments}
\pst The authors thank the Russian Foundation for Basic Research
for partial support under Projects Nos. 09-02-00142 and
10-02-00312. S.N.F. thanks the Ministry of Education and Science
of the Russian Federation and the Federal Education Agency for
support under Project No. 2.1.1/5909.


\begin{thebibliography}{99}

\bibitem{Rita1}
M. A. Man'ko, {\sl J. Russ. Laser Res.}, {\bf 21}, 411 (2000).

\bibitem{Rita1a}
M. A. Man'ko, {\sl J. Russ. Laser Res.}, {\bf 22}, 48 (2001).

\bibitem{Rita1b}
M. A. Man'ko, {\sl J. Russ. Laser Res.}, {\bf 23}, 433 (2002).

\bibitem{Rita2}
M. A. Man'ko, {\sl J. Russ. Laser Res.}, {\bf 27}, 405 (2006).

\bibitem{Rita2a}
M. A. Man'ko, {\sl J. Russ. Laser Res.}, {\bf 22}, 168 (2001).

\bibitem{Rita3}
R. Fedele and M. A. Man'ko, {\sl Eur. Phys. J. D}, {\bf 27}, 263 (2003).

\bibitem{Rita3a}
M. A. Man'ko, O. V. Man'ko, and V. I. Man'ko, {\sl Int. J. Mod. Phys. D},
{\bf 20}, 1399 (2006).

\bibitem{Rita3b}
M. A. Man'ko, {\sl J. Russ. Laser Res.}, {\bf 27}, 507 (2006).

\bibitem{Rita3c}
M. A. Man'ko, {\sl J. Russ. Laser Res.}, {\bf 30}, 514 (2009).

\bibitem{FP}
M. A. Man'ko and V. I. Man'ko, {\sl Found. Phys.}, DOI
10.1007/s10701-009-9403-9 (2009).

\bibitem{tombesi-manko}
S. Mancini, V. I. Man'ko, and P. Tombesi, {\sl Phys. Lett. A}, {\bf 213}, 1 (1996).

\bibitem{m-t-m-FondPhys}
S. Mancini, V. I. Man'ko, and P. Tombesi, {\sl Found. Phys.}, {\bf 27}, 801 (1997).

\bibitem{schrodinger}
E. Schr\"{o}dinger, {\sl Ann. Phys., Lpz.}, {\bf 79}, 489 (1926).

\bibitem{landau}
 L. D. Landau, {\sl Ztschr. Physik}, {\bf 45}, 430 (1927).

\bibitem{von-neumann}
J. von Neumann, {\sl G\"{o}ttingen.\ Nachr.}, 245 (1927).

\bibitem{wigner}
E. P. Wigner, {\sl Phys. Rev.}, {\bf 40}, 749 (1932).

\bibitem{husimi}
K. Husimi, {\sl Proc. Phys. Math. Soc. Jpn}, {\bf 22}, 264 (1940).

\bibitem{sudarshan}
E. C. G. Sudarshan, {\sl Phys. Rev. Lett.}, {\bf 10}, 177 (1963).

\bibitem{glauber}
R. J. Glauber, {\sl Phys. Rev. Lett.}, {\bf 10}, 84 (1963).

\bibitem{radon}
J. Radon, {\sl Ber. Verh. Saechs. Akad. Wiss. Leipzig,
Math.-Phys. Kl.}, {\bf 69}, 262 (1917).

\bibitem{gelfand}
I. M. Gel'fand and G. E. Shilov, {\it Generalized Functions: Properties
and Operations}, Academic Press, New York (1966), Vol.~5.

\bibitem{dodonovPLA}
V. V. Dodonov and V. I. Man'ko, {\sl Phys. Lett. A}, {\bf 229}, 335 (1997).

\bibitem{oman'ko-jetp}
V. I. Man'ko and O. V. Man'ko, {\sl J. Exp. Theor. Phys.}, {\bf 85}, 430 (1997).

\bibitem{serg-spin}
S. N. Filippov and V. I. Man'ko, {\sl J. Russ. Laser
Res.}, {\bf 30}, 129 (2009).

\bibitem{serg-inverse-spin}
S. N. Filippov and V. I. Man'ko, {\sl J. Russ. Laser
Res.}, {\bf 31}, 32 (2010).

\bibitem{serg-distances}
S. N. Filippov and V. I. Man'ko, ``Distances between quantum
states in the tomographic-probability representation,"
arXiv:0911.1414v1 [quant-ph] (2009).

\bibitem{ibort}
A. Ibort, V. I. Man'ko, G. Marmo, et al.,
{\sl Phys. Scr.}, {\bf 79}, 065013 (2009).

\bibitem{berber}
J. Bertrand and P. Bertrand, {\sl Found. Phys.}, {\bf 17}, 397 (1987).

\bibitem{vogel}
K. Vogel and H. Risken, {\sl Phys. Rev. A}, {\bf 40}, 2847 (1989).

\bibitem{mancini95}
S. Mancini, V. I. Man'ko, and P. Tombesi, {\sl Quantum Semiclass. Opt.}, {\bf 7}, 615
(1995).

\bibitem{banaszek}
K. Banaszek and K. Wodkiewicz, {\sl Phys. Rev. Lett.}, {\bf 76}, 4344 (1996).

\bibitem{wvogel}
S. Wallentowitz and W. Vogel, {\sl Phys. Rev. A}, {\bf 53}, 4528 (1996).

\bibitem{mancini}
S. Mancini, P. Tombesi, and V. I. Man'ko, {\sl Europhys. Lett.}, {\bf 37}, 79 (1997).

\bibitem{asorey}
M. Asorey, P. Facchi, V. I. Man'ko, et al., {\sl Phys. Rev. A}, {\bf 77},
042115 (2008).

\bibitem{asorey-arxiv}
M. Asorey, P. Facchi, G. Florio, et al, ``Robustness of raw
quantum tomography," arXiv:1003.1664v1 [quant-ph] (2010).

\bibitem{ali-prugovecki-1}
S. T. Ali and E. Prugove\v{c}ki, {\sl J. Math.
Phys.}, {\bf 18}, 219 (1977).


\bibitem{ali-prugovecki-2}
S. T. Ali and E. Prugove\v{c}ki, {\sl Physica A}, {\bf 89}, 501 (1977).

\bibitem{ali-prugovecki-3}
S. T. Ali and E. Prugove\v{c}ki, {\sl Int. J. Theor. Phys.}, {\bf 16}, 689 (1977).

\bibitem{bush-lahti-FounPhys}
P. Busch and P. J. Lahti, {\sl Found. Phys.}, {\bf 19},
633 (1989).

\bibitem{bush-lahti}
P. Busch, G. Cassinelli, and P. J. Lahti, {\sl Rev. Math. Phys.}, {\bf 7}, 1105 (1995).

\bibitem{bush-lahti-book}
P. Busch, M. Grabowski, and P. J. Lahti, {\it Operational Quantum
Physics}, {\sl Lecture Notes in Physics}, Springer-Verlag,
Berlin (1995), Vol.~31.

\bibitem{stulpe}
W. Stulpe, {\sl Found. Phys.}, {\bf 24}, 1089 (1994).

\bibitem{stulpe-article}
W. Stulpe, {\sl Int. J. Theor. Phys.}, {\bf 37}, 349 (1998).

\bibitem{stulpe-book}
W. Stulpe, ``Classical representations of quantum
mechanics related to statistically complete observables," Los
Alamos Arxiv, quant-ph/0610122 (2006).

\bibitem{kiukas}
J. Kiukas, P. Lahti, and J.-P. Pellonp\"{a}\"{a},
{\sl J. Phys. A: Math. Theor.},  {\bf 41}, 175206 (2008).

\bibitem{amiet-weigert-JPA}
J.-P. Amiet and S. Weigert, {\sl J. Phys. A: Math. Gen.}, {\bf 32}, L269 (1999).

\bibitem{newton}
R. G. Newton and B. Young, {\sl Ann. Phys.}, {\bf 49}, 393 (1968).

\bibitem{caves-sic}
C. M. Caves, ``Symmetric informationally complete POVMs," UNM
Information Physics Group internal report,
http://info.phys.unm.edu/~caves/reports/infopovm.pdf (1999).

\bibitem{caves}
C. M. Caves, C. A. Fuchs, and R. Schack, {\sl Phys. Rev.
A}, {\bf 65}, 022305 (2002).

\bibitem{renes}
J. M. Renes, R. Blume-Kohout, A. J. Scott, and C. M. Caves,
{\sl J. Math. Phys.}, {\bf 45}, 2171 (2004).

\bibitem{fuchs-2010}
D. M. Appleby, S. T. Flammia, and C. A. Fuchs, ``The Lie algebraic
significance of symmetric informationally complete measurements,"
arXiv:1001.0004v1 [quant-ph] (2010).

\bibitem{fuchs-perimeter}
C. A. Fuchs, ``Quantum Bayesianism at the Perimeter,"
arXiv:1003.5182v1 [quant-ph] (2010).

\bibitem{stratonovich}
R. L. Stratonovich, {\sl Sov. Phys. JETP}, {\bf 4}, 891 (1957).

\bibitem{oman'ko-JPA}
O. V. Man'ko, V. I. Man'ko, and G. Marmo, {\sl J. Phys. A: Math.
Gen.}, {\bf 35}, 699 (2002).

\bibitem{oman'ko-vitale}
O. V. Man'ko, V. I. Man'ko, G. Marmo, and P. Vitale,
{\sl Phys. Lett. A}, {\bf 360}, 522 (2007).

\bibitem{serg-chebyshev}
S. N. Filippov and V. I. Man'ko, {\sl J. Russ. Laser
Res.}, {\bf 30}, 224 (2009).

\bibitem{serg-PhysScr}
S. N. Filippov and V. I. Man'ko, {\sl Phys.
Scr.}, {\bf 79}, 055007 (2009).

\bibitem{nikiforov-suslov-uvarov}
A. F. Nikiforov, S. K. Suslov, and V. B. Uvarov, {\it Classical
Orthogonal Polynomials of a Discrete Variable}, Springer-Verlag,
Berlin, Heidelberg, New York (1991).

\bibitem{man'ko-sudarshan}
V. I. Man'ko, G. Marmo, E. C. G. Sudarshan, and F. Zaccaria,
{\it Phys. Lett. A}, {\bf 327}, 353 (2004).

\bibitem{scott-grassl}
A. J. Scott and M. Grassl, {\sl J. Math. Phys.}, {\bf 51}, 042203
(2010).

\bibitem{ericsson}
D. M. Appleby, \AA. Ericsson and C. A. Fuchs, {\sl Found. Phys.},
DOI: 10.1007/s10701-010-9458-7 (2010).

\bibitem{weigert-PRL}
S. Weigert, {\sl Phys. Rev. Lett.}, {\bf 84}, 802 (2000).

\bibitem{manko-sudarshan-vent}
V. I. Man'ko, G. Marmo, A. Simoni, et al., {\sl Rep. Math.
Phys.}, {\bf 61}, 337 (2008).

\bibitem{jones-linden}
N. S. Jones and N. Linden, {\sl Phys. Rev. A}, {\bf 71}, 012324
(2005).

\bibitem{gantmacher}
F. R. Gantmacher, {\it The Theory of Matrices}, AMS Chelsea Publishing, Providence, RI (1998).

\bibitem{lemmens}
P. W. H. Lemmens and J. J. Seidel, {\sl J.
Algebra}, {\bf 24}, 494 (1973).

\bibitem{tremain}
J. C. Tremain, ``Concrete constructions of real equiangular line
sets," arXiv:0811.2779v1 [math.MG] (2008).

\bibitem{filipp-chebyshev}
S. N. Filippov and V. I. Man'ko, {\sl J. Russ. Laser Res.}, \textbf{30}, 224 (2009).

\bibitem{albouy}
O. Albouy and M. R. Kibler, {\sl J. Russ. Laser Res.}, {\bf
28}, 429 (2007).

\bibitem{appleby-MUBs}
D. M. Appleby, ``SIC-POVMs and MUBs: geometrical relationships in
prime dimension," arXiv:0905.1428v1 [quant-ph] (2009).

\bibitem{gelfand-minlos-shapiro}
I. M. Gel'fand, R. A. Minlos, and Z. Ya. Shapiro,
{\it Representations of the Rotation and Lorentz Groups and Their
Applications}, The Macmillan Company, New York (1963).

\bibitem{smorodinskii}
Ya. A. Smorodinskii and M. Huszar, {\sl Theor. Math. Phys.},
{\bf 4}, 867 (1970).

\end{thebibliography}
\end{document}